\documentclass[preprint]{aastex}

\received{}
\accepted{}
\journalid{}{}
\articleid{}{}
\slugcomment{}

\newcommand{\etal}{{\em et~al.}}
\newcommand{\ug}{$u'\!-\!g'$}
\newcommand{\gr}{$g'\!-\!r'$}
\newcommand{\ri}{$r'\!-\!i'$}
\newcommand{\iz}{$i'\!-\!z'$}
\newcommand{\ugriz}{$u',g',r',i',z'$}

\newcommand{\p}[1]{{\tt #1}}
\newcommand{\myurl}[1]{{\em\underline{#1}}}

\shortauthors{Csabai et~al.}
\shorttitle{Photometric Redshifts for the SDSS Early Data Release}

\begin{document}

\title{The Application of Photometric Redshifts to the SDSS Early Data Release}

\author{
Istv\'an Csabai\altaffilmark{\ref{Eotvos},\ref{JHU}},
Tam\'as  Budav\'ari\altaffilmark{\ref{JHU}}, 
Andrew J. Connolly\altaffilmark{\ref{Pittsburgh}},
Alexander S. Szalay\altaffilmark{\ref{JHU},\ref{Eotvos}},
Zsuzsanna Gy\H ory\altaffilmark{\ref{Eotvos}},
Narciso Ben\'{\i}tez\altaffilmark{\ref{JHU}},
Jim Annis\altaffilmark{\ref{Fermilab}},
Jon Brinkmann\altaffilmark{\ref{APO}},
Daniel Eisenstein\altaffilmark{\ref{Arizona}},
Masataka Fukugita\altaffilmark{\ref{CosmicRay}},
Jim Gunn\altaffilmark{\ref{Princeton}},
Stephen Kent\altaffilmark{\ref{Fermilab}},
Robert Lupton\altaffilmark{\ref{Princeton}},
Robert C. Nichol\altaffilmark{\ref{CMU}},
Chris Stoughton\altaffilmark{\ref{Fermilab}}
%et~al.
}

\altaffiltext{1}{Department of Physics, E\"{o}tv\"{o}s University,
Budapest, Pf.\ 32, Hungary, H-1518
\label{Eotvos}}

\altaffiltext{2}{Department of Physics and Astronomy, The Johns Hopkins University, 3701 San Martin Drive, Baltimore, MD~21218
\label{JHU}}

\altaffiltext{3}{Department of Physics and Astronomy, University of
Pittsburgh, Pittsburgh, PA 15260
\label{Pittsburgh}}

%\altaffiltext{4}{Astrophysikalisches Institut Potsdam, An der
%Sternwarte 16, D-14482 Potsdam, Germany 
%\label{AIP}}

\altaffiltext{4}{Fermi National Accelerator Laboratory, P.O. Box 500,
Batavia, IL~60510
\label{Fermilab}}

%\altaffiltext{6}{Department of Astronomy and Astrophysics,
%The Pennsylvania State University, University Park, PA 16802
%\label{PennState}}

\altaffiltext{5}{Apache Point Observatory, P.O. Box 59,
Sunspot, NM 88349-0059
\label{APO}}

%\altaffiltext{8}{Department of Astronomy, California Institute of
%Technology, Pasadena, CA 91125 
%\label{Caltech}}

\altaffiltext{6}{Princeton University Observatory, Princeton, NJ~08544
\label{Princeton}}

%\altaffiltext{10}{Pontificia Universidad Cat\'{o}lica de Chile,
%Departamento de Astronom\'{\i}a y Astrof\'{\i}sica,
%Facultad de F\'{\i}sica, Casilla 306, Santiago 22, Chile
%\label{Chile}}

%\altaffiltext{11}{U.S. Naval Observatory, 
%3450 Massachusetts Ave., NW, 
%Washington, DC  20392-5420
%\label{USNO}}

\altaffiltext{7}{Steward Observatory, 933 N. Cherry Ave, Tucson, AZ 85721
\label{Arizona}}

\altaffiltext{8}{Institute for Cosmic Ray Research, University
of Tokyo, Midori, Tanashi, Tokyo 188-8502, Japan
\label{CosmicRay}}

\altaffiltext{9}{Dept. of Physics, Carnegie Mellon University,
5000 Forbes Ave., Pittsburgh, PA~15232
\label{CMU}}

%\altaffiltext{13}{University of Chicago, Astronomy \& Astrophysics
%Center, 5640 S. Ellis Ave., Chicago, IL 60637
%\label{Chicago}}

\begin{abstract}
The Early Data Release from the Sloan Digital Sky survey provides one
of the largest multicolor photometric catalogs currently available to
the astronomical community. In this paper we present the first
application of photometric redshifts to the $\sim 6$ million extended
sources within these data (with 1.8 million sources having $r' < 21$). 
Utilizing a range of photometric redshift techniques, from
empirical to template and hybrid techniques, we investigate the
statistical and systematic uncertainties present within the redshift
estimates for the EDR data. For $r'<21$ we find that the redshift
estimates provide realistic redshift histograms with an rms
uncertainty in the photometric redshift relation of 0.035 at $r'<18$
and rising to 0.1 at $r'<21$. We conclude by describing how these
photometric redshifts and derived quantities, such as spectral type,
restframe colors and absolute magnitudes, are stored within the SDSS
database. We provide sample queries for searching on photometric
redshifts and list the current caveats and issues that should be
understood before using these photometric redshifts in statistical
analyses of the SDSS galaxies.

\end{abstract}


\section{Introduction} 
\label{sec:intro}

From their inception (\citet{koo85, connolly95a, gwyn96, sawicki97, hogg98,
wang98, soto99, benitez00, csabai00, budavari00})
photometric redshifts have been seen as an efficient and effective means of
studying the statistical properties of galaxies and their evolution. They are
essentially a mechanism for inverting a set of observable parameters (e.g.\
colors) into estimates of the physical properties of galaxies (e.g.\
redshift, type and luminosity). To date photometric redshifts have typically
been employed on small multicolor photometric surveys such as the Hubble Deep
Field (HDF, \citet{williams96}). While these applications have demonstrated the power of
the estimated redshifts in studying galaxy evolution they have an underlying
limitation. The cosmological volumes probed by the narrow pencil beam surveys
are small and consequently it is not clear if these data provide a
representative sample of the Universe. With the development of large
wide-field survey cameras this volume limitation can be overcome and large,
statistically complete studies of the properties of galaxies can be
undertaken.

One of the largest ongoing multicolor photometric survey currently
underway is the Sloan Digital Sky Survey (\citet{york}). This imaging
and spectroscopic survey provides an ideal base from which to apply
photometric redshifts to large samples of galaxies. In the Early Data
Release (\citet{stoughton02}) there are over 6 million galaxies, an order of
magnitude increase in sample size when compared to existing public
multicolor surveys.  From these galaxies there are approximately 35,000 galaxies with
published spectroscopic redshifts from which to determine the
statistical and systematic uncertainties within the SDSS photometric
redshift relation.

In this paper we describe the first application of photometric
redshifts to the SDSS data. We provide a background to the redshift
estimation techniques but do not go into the technical details of the
individual methods. We focus on
providing the astronomical community with details of how to use the
photometric redshifts within the SDSS EDR database and emphasize the
caveats and limitations present within the current photometric
redshift catalog (due to photometric errors and uncertainties in the
SDSS zeropoints). We plan to have a more detailed analysis of systematic
errors on the soon outcoming Data Release 1, where most of these
problems will be eliminated. Sample queries for the EDR database are provided
in Section 6.1 together with details of value added parameters that can be
derived from the photometric redshifts such as restframe colors,
k-corrections and absolute magnitudes.

\section{The Early SDSS Data Release} 
\label{sec:data}

\begin{figure}
%\plotone{figs/zhist1.eps}
\plotone{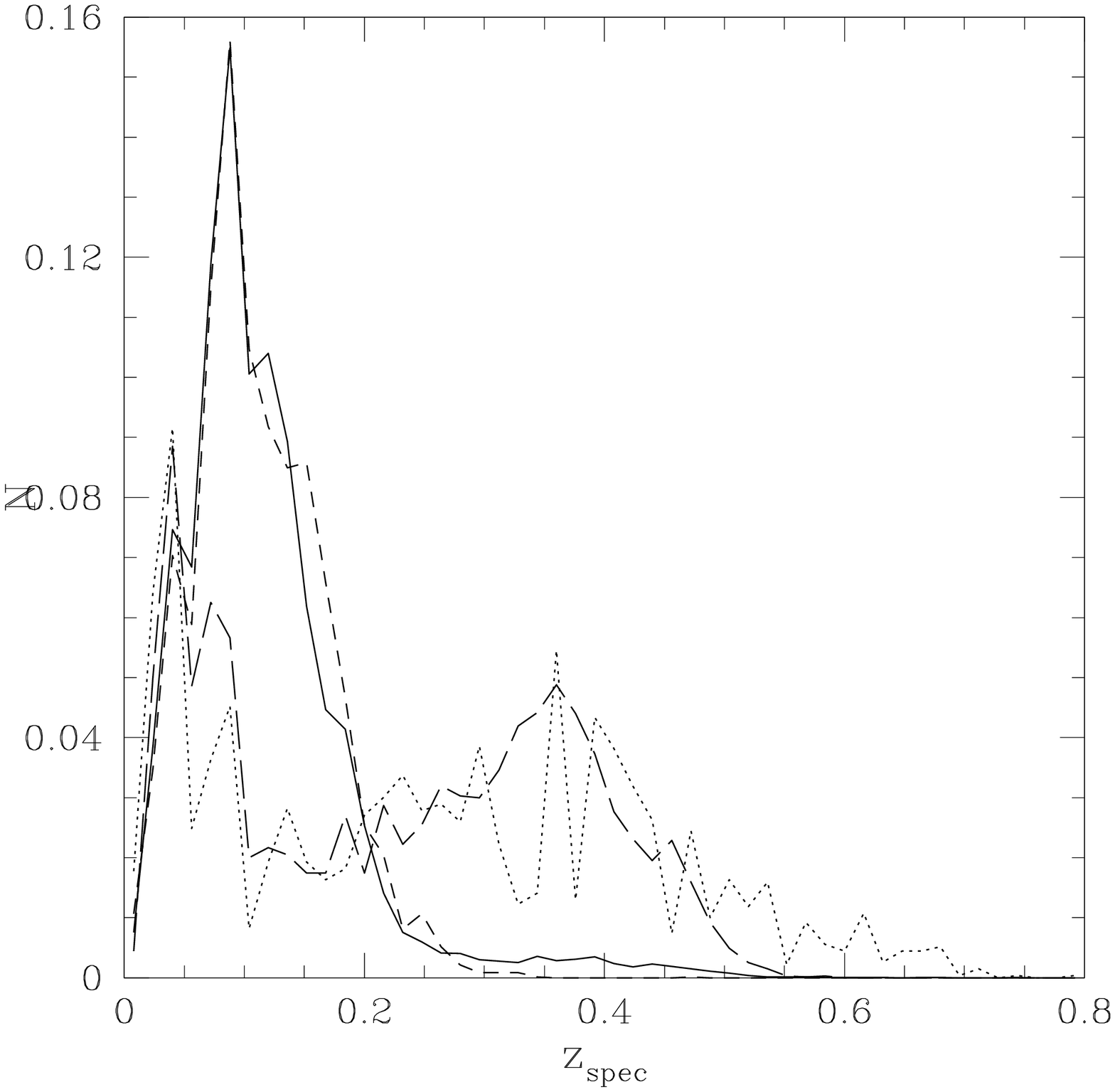}
\caption[]{The spectroscopic redshift histogram for the SDSS main EDR (solid), the EDR
LRG (long dash), the 2dF (short dash) and the CNOC2 sets.   }
\label{fig:zhist1}
\end{figure}

In this section we provide a brief description of the Early Data
Release (EDR; \citet{stoughton02}) of the SDSS and introduce the
subsets of the data that will be used throughout this paper. The EDR
has 5 band photometry \citep{fukugita, gunn, smith, hogg, pier} for over 
6 million galaxies. Out of which 1.8
million galaxies have $r' < 21$. The five filters of the \ugriz system
have effective wavelengths of 3540, 4750, 6222, 7632, and 9049\AA
respectively, and the goal of the survey is to achieve a level of
photometric uniformity and accuracy such that the
systemwide rms errors in the final SDSS photometric catalog will be less
than 0.02 mag in  $r'$, 0.02 mag in \ri and \gr, and 0.03 mag in \ug and \iz, 
for objects bluer than an M0 dwarf. 
All analyses in this paper are based
on the dereddened model magnitudes in the EDR data set. A relatively
small subset of these galaxies, $>$30,000, have measured
redshifts. The objects for spectroscopic observation were selected
using the SDSS's target selection algorithm, which is discussed in
detail in \citet{stoughton02} and \citet{strauss02}. This selection
algorithm results in two subsets of the SDSS data, a main galaxy
sample and a luminous red galaxy sample
(LRG; \citet{eisenstein01}). The main galaxy sample contains 27,797
galaxies with a mean redshift of $z=0.116$ and a photometric limit of
$r'=18$. The LRG sample was selected from galaxies with colors
similar to that of an elliptical galaxy and contains $6698$ galaxies
with a mean redshift of $z=0.227$ (though extending out to
$z>0.5$). The redshift histograms of these two subsets of the data are
given in Figure~\ref{fig:zhist1} which demonstrate that the main
sample should provide a good training/test set out to $z=0.2$ and the
LRG data set out to $z=0.5$.

In order to test the accuracy of the photometric redshifts derived
from the SDSS we supplement the SDSS redshifts with a subset of
galaxies selected from published redshift catalogs. At low redshift
and for bright magnitudes, the 2 degree Field (2dF) redshift survey
(\citet{colless01}) contains $5642$ galaxies for which we have matching
SDSS photometry. These galaxies have a limiting magnitude of
approximately $r'=18.5$ and a mean redshift of $z=0.112$. The
redshift range sampled by these galaxies is, therefore, well matched
to that of the SDSS redshift catalog with a limiting redshift of
approximately $z=0.2$. At higher redshifts and for fainter magnitudes,
the Canada Network for Observational Cosmology (CNOC2; \citet{yee96})
survey has magnitude limit of approximately $r'<21.0$ with a mean
redshift of $z=0.274$ and an upper redshift limits of approximately
$z=0.7$. The photometric depth of the $2697$ galaxies within the the
CNOC2 sample provides not just a test of the accuracy of the
photometric redshifts but also a measure of how the redshift
uncertainties scale with magnitude limit. We designate these ``blind''
test samples as {\it 2dF} for the low redshift samples and CNOC2 for
the CNOC2 data.

In the following sections we will use the main EDR and the EDR LRG
samples as training sets and all of the above data sets as test sets.

\section{Standard Photometric Redshift Techniques}

A wide range of techniques have been employed in the literature to
estimate redshifts of galaxies with broadband photometric colors. Approaches have ranged from the purely empirical relations to
comparisons of the colors of galaxies to the colors predicted from
galaxy spectral energy distributions. Each approach has its own set of
advantages and disadvantages. Empirical approaches, where the
color-redshift relations are derived directly from the data
themselves, are relatively free from possible systematic effects
within the photometric calibration. As such, they provide a simple
measure of the statistical uncertainties with the data and can
demonstrate the accuracy to which we should be able to estimate
redshifts once we can control the systematic errors. Their underlying
disadvantage is that we can typically only apply these relations to
galaxies with colors that lie within the range of colors and redshifts
found within the training set. Template based techniques are free from
the limitation of a training set and can be applied over a wide range
of redshifts and intrinsic colors. They rely, however, on having a
set of galaxy templates that accurately map the true distribution of
galaxy spectral energy distributions (and their evolution with
redshift) and on the assumption that the photometric calibration of
the data is free from systematics.

In this section we consider both empirical and template based
approaches to photometric redshift estimation for SDSS data. We
demonstrate the redshift accuracy that it should be possible to
achieve from the EDR sample and describe the current limitations of
using standard galaxy spectral energy distributions.

\subsection{Empirical Redshift Estimation Methods} 
\label{sec:emp}

\begin{figure}
%\plottwo{figs/zzNnn.eps}{figs/zzPoly.eps} % {figs/zzKdtree.eps}{}
\plottwo{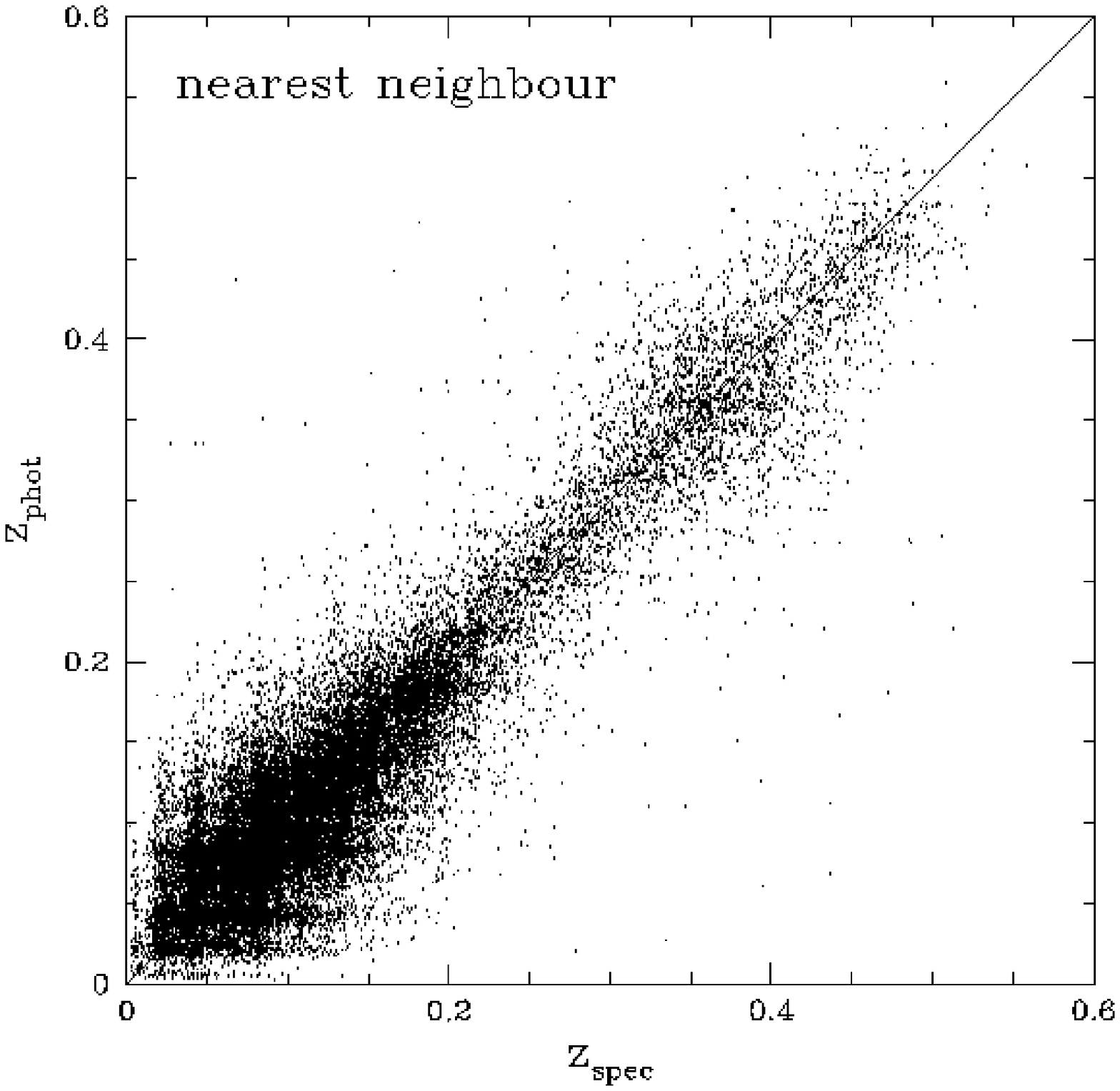}{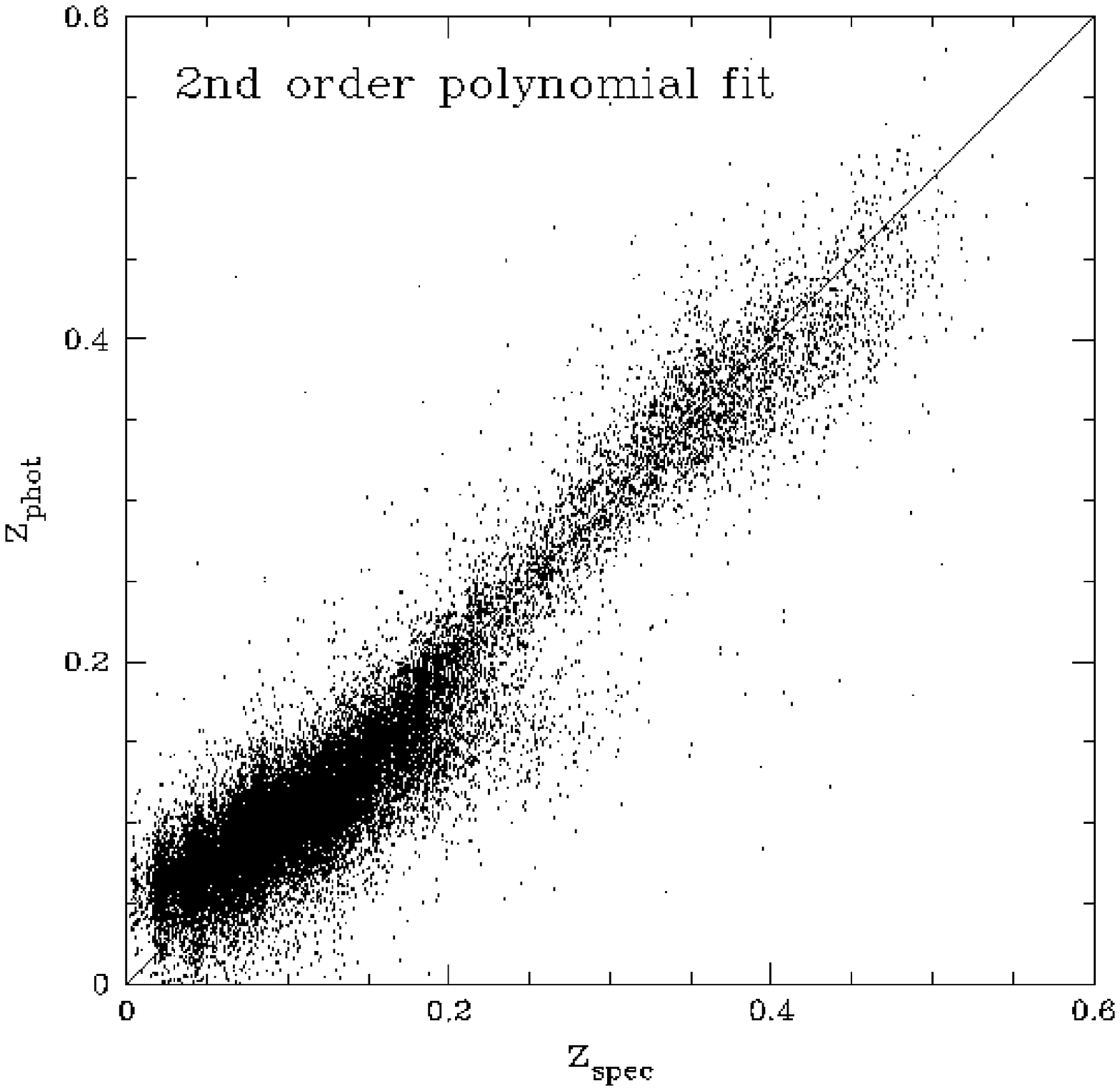}
\caption[]{The photometric redshift estimations with the simple
empirical methods.}
\label{fig:emp1.figzz}
\end{figure}

We consider here the standard empirical redshift estimation techniques
that have been used in the literature (\citet{connolly95a, wang98,
brunner99}) and develop a new technique based on a hierarchical
indexing structures (kd-trees,\citet{moore}). One of the first successful
empirical methods is based on fitting a functional form for the
relation between the spectroscopic redshift of a galaxy and its colors
or magnitudes (\citet{connolly95a}).  This function is typically a 2nd
or 3rd order polynomial.  Figure~\ref{fig:emp1.figzz} shows the
photometric vs. the spectroscopic redshifts using the EDR main galaxy
and LRG spectroscopic samples. As the size of the training set is
large ($>30,000$) when compared to the number of the fitted parameters
($21$), we can expect that this fit will work for other objects with
the same dispersion as seen in Figure~\ref{fig:emp1.figzz} (as long as
the data are selected over the sample color and redshift range as the
training set). The dispersion within this photometric redshift
relation is $\sigma_z = 0.027$ (see Table~\ref{tab:rms.tbl} for
comparision with other values).  One possible uncertainty within this
technique comes from the fact that the fitting function is just an
approximation of the, possibly, more complex relation between the
colors and the redshift of a galaxy. We would, therefore, expect the
fitting function to accurately follow the redshift-color relation over
a narrow range of redshift. A technique to avoid this, is to use
separate functions in different redshift (\citet{brunner99}) or color
ranges.

\begin{figure}
\plotone{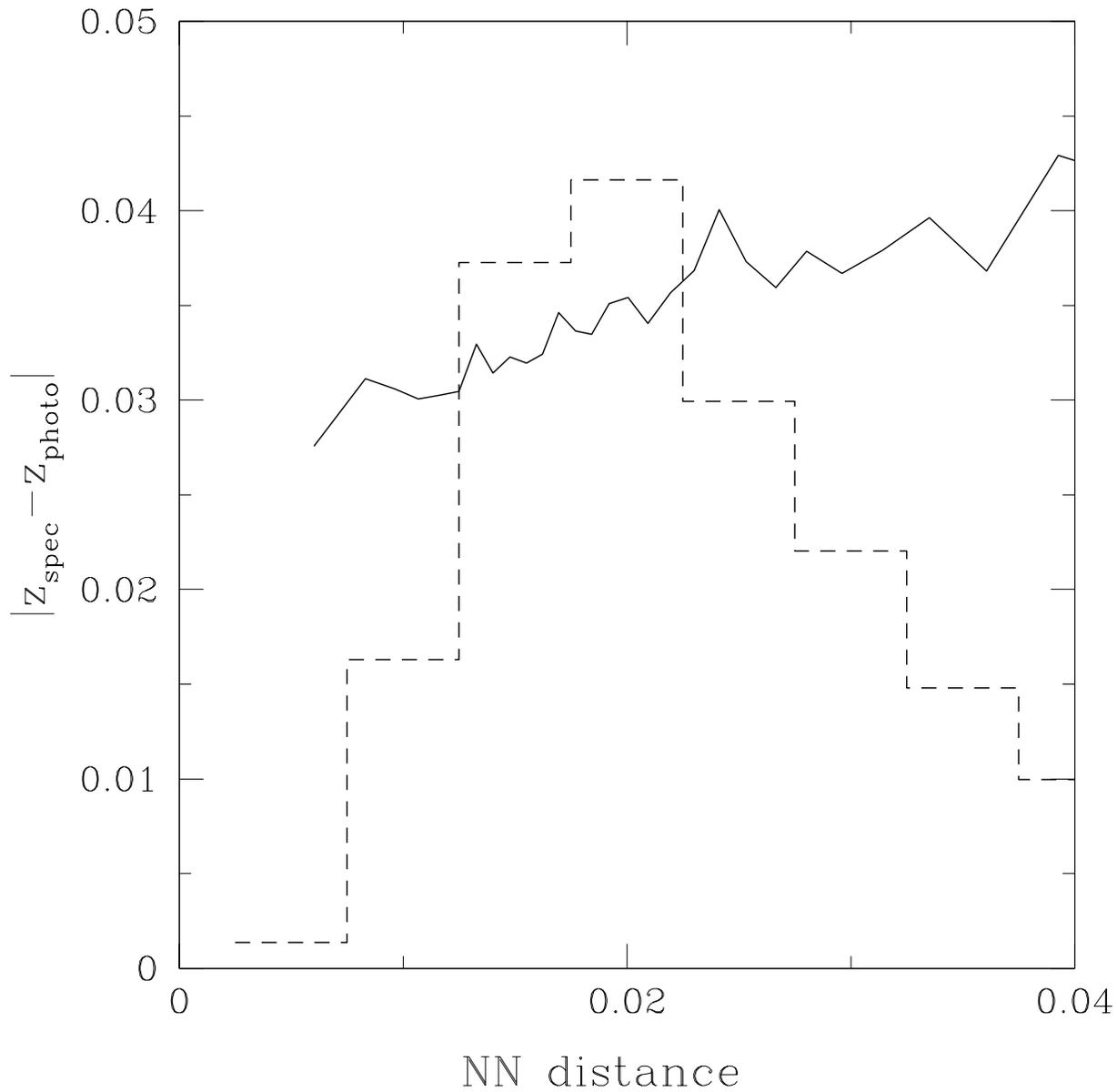}
\caption[]{The dependence of redshift average estimation error on the color
space distance from the nearest reference object (solid line). As
expected, smaller distances result smaller error.  
The dashed line is for the histogram of number of objects with a given
nearest neighbor distance. One can see, that for most of the objects the
nearest neighbor is not close enough to get the best estimation.
}
\label{fig:nn}
\end{figure}

A second, and possibly the simplest, empirical estimator is the
nearest neighbor method. For a test galaxy, this finds the galaxy
within the training set with the smallest distance in the color (or
magnitude) space (weighted by the errors). The redshift of this
closest match is then assigned to the test galaxy. In the ideal case
the training set contains sufficient galaxies that for each unknown
object there is a close neighbor. In Figure~\ref{fig:nn} we show,
that redshift estimation error increases with the distance from the    
nearest neighbor in color space. The larger the dataset the more
accurate this method becomes, as long as that all galaxy types are
represented in the training set. 
From the technical viewpoint, larger
training sets mean that the search time increases so one has to use an
efficient multidimensional search technique (e.g.\ kd-trees) instead
of a standard linear search. The comparison between the estimated and
spectroscopic redshifts for the nearest-neighbor technique is given in
Figure~\ref{fig:emp1.figzz}. The dispersion about this relation is
$\sigma_z = 0.033$.

\begin{figure}
%\plottwo{figs/zzKdtree.eps}{figs/kdtreeBox.eps}
\plottwo{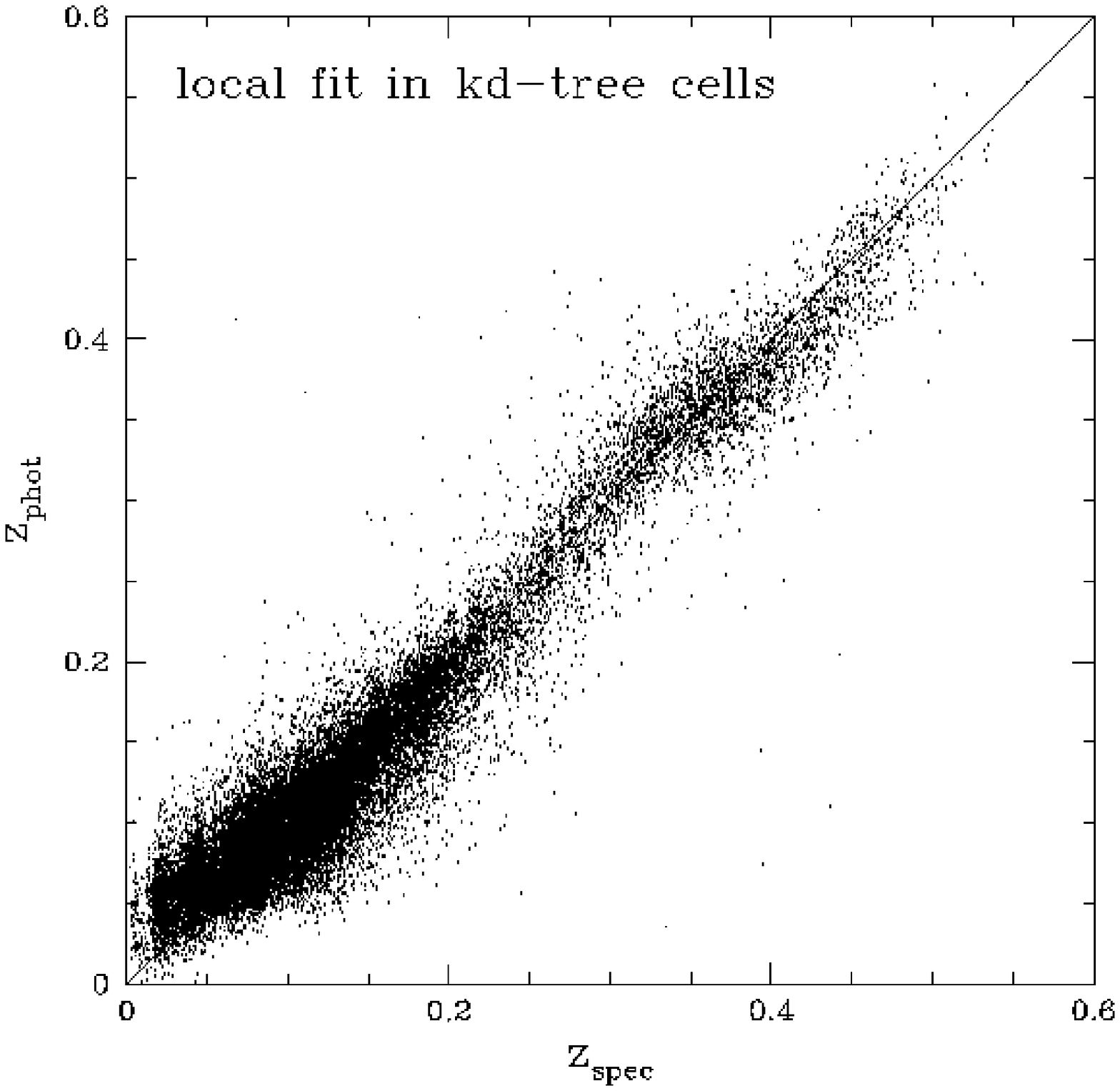}{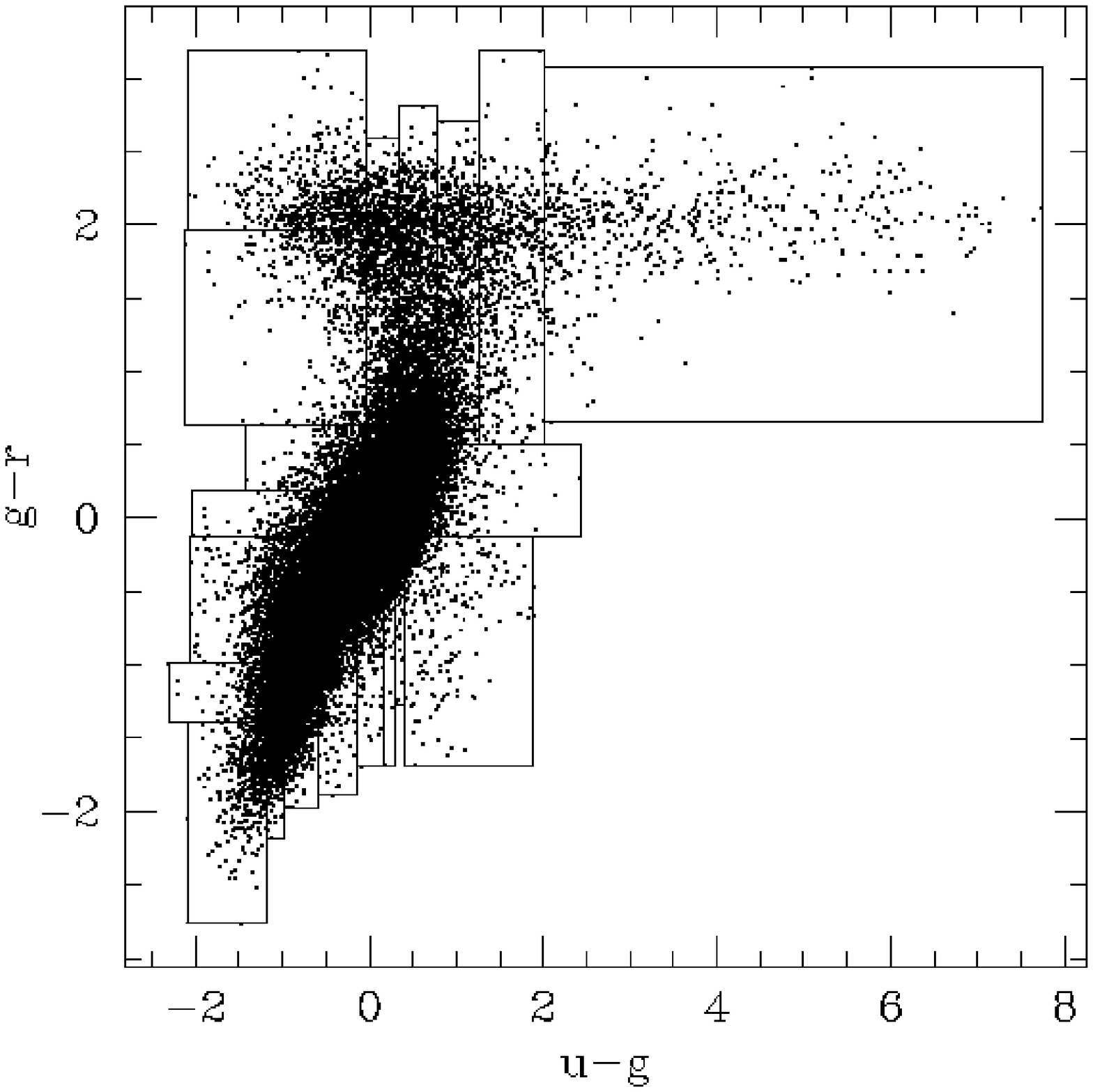}
\caption[]{On the right we plot a 2 dimensional demonstration of the
color space partitioning. In each of these cells we applied the
polynomial fitting technique to estimate redshifts. The left figure show
the results.}
\label{fig:emp2.figzz}
\end{figure}

A natural limitation of the nearest neighbor technique is that a
large number of training galaxies alone is not enough, they must cover
the range of the colors of the unknown objects in a more or less
uniform way. Unfortunately, this is usually not the case. To resolve
this problem one can search for more than one nearest neighbor and
apply an interpolation or a fitting function. This also helps to resolve
a second problem, namely that because of the finite number of objects
in the training set, the photometric redshifts will have discrete
values making them problematic to use in some statistical studies. We
have created a hybrid version of the above two empirical methods: we
partitioned the color space into cells, containing the same number of
objects from the training set, using a kd-tree tree (a binary search
tree \citep{bentley79}).  In each cell we
fit a second order polynomial. The results together with a
demonstration of a 2-dimensional version of the kd-tree partitioning
of the EDR training set are given in Figure~\ref{fig:emp2.figzz}. The
dispersion about this relation is $\sigma_z = 0.023$.

For each of these approaches the resulting dispersion in the
photometric redshift relation is found to be approximately 0.03 (see
Table~\ref{tab:rms.tbl} with the hybrid method being marginally more
accurate. As these empirical approaches do not rely on the absolute
photometric calibration of the data (other than the calibration should
be stable across the data sets) they are somewhat insensitive to
systematic errors in the data. If the SDSS redshifts (or external
redshift samples) sampled the full redshift range of the data to the
limit of the survey these empirical techniques would provide an ideal
mechanism for deriving redshift estimates for the SDSS. As the
redshift range of the spectroscopic samples are fairly limited the
application of these techniques to the full data set is
non-trivial. We can, however, use these results to demonstrate that
accuracy we should be able to derive from the template based
techniques (once any systematics within the data are accounted for)
should be $\sigma_z \sim 0.03$ at $r'<18$.

\subsection{Template Based Redshift Estimation Methods} 
\label{sec:sedfit}

As noted previously, the advantage of using templates to estimate
redshifts of galaxies (\citet{koo85, gwyn96, sawicki97, connolly99,
soto99, benitez00, bolzonella00, budavari99, budavari00, csabai00}) are
numerous. This approach simply compares the expected colors of a
galaxy (derived from template spectral energy distributions) with
those observed for an individual galaxy. The standard scenario for
template fitting is to take a small number of spectral templates $T$
(e.g.\ E, Sbc, Scd and Irr galaxies) and choose the best fit by
optimizing the likelihood of the fit as a function of redshift, type
and luminosity $p(z,T,\cal{L})$. Variations on this approach have been
developed in the last few decades including ones that use a continuous
distribution of spectral templates enabling the error function in
redshift and type to be well defined.

A representative set of spectrophotometrically calibrated spectral
templates is not easy to obtain. One problem with measured spectra is,
that to calibrate them spectrophotometrically over the full spectral
range is non-trivial. A second problem is that, because of the
redshift of a galaxy, we need spectra over a wavelength range that is
wider than the range of our optical filters (3000--12000\AA). Such
spectra cannot currently be measured by a single spectrograph. Third,
even if we could measure calibrated spectra over the required range,
spectrographs, especially modern multifiber ones, usually sample only
the central region of the galaxy while photometric measurements
integrate over the full spatial extent of a galaxy. The alternative to
empirical templates is to use the outputs of spectral synthesis
models. The accuracy of spectral models are improving \citep{bc93} but
not yet as accurate as direct measurements of galaxy spectra. Modern
surveys will improve on this situation, e.g.\ the SDSS will measure
spectrophotometrically calibrated spectra for a million objects in the
3800--9200\AA{} range at a resolution $R=\lambda / \Delta\lambda$ of
about 1800, but to-date there does not exist an optimal set of galaxy
spectral templates.

\begin{figure}
%\plotone{figs/zzCww.eps}
\plotone{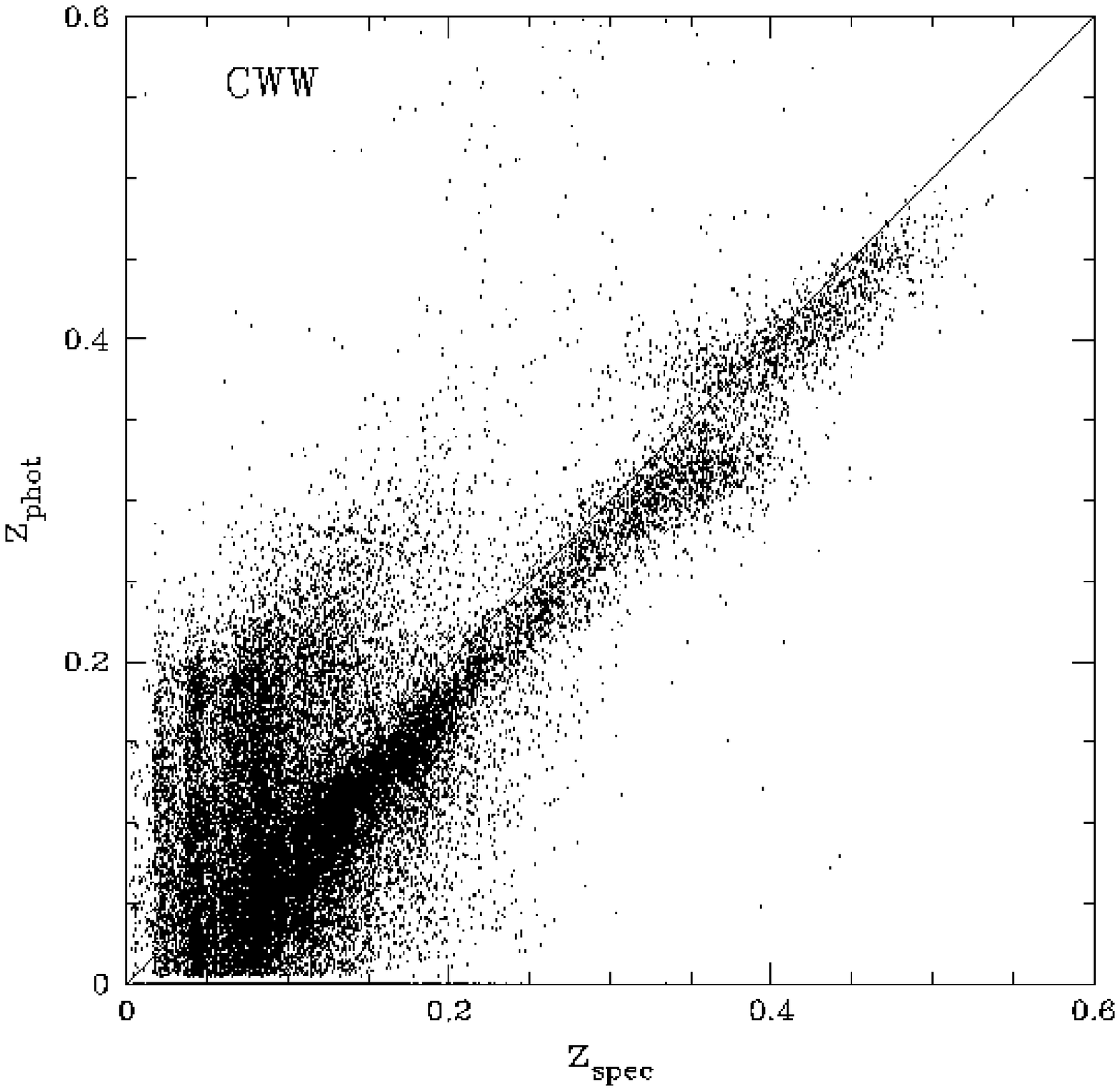}
\caption[]{Photometric redshift estimation using the CWW spectral
energy distributions. The rms dispersion about this relations is 0.062
in redshift. For the majority of the SDSS galaxies the CWW templates
perform reasonably well (with the core of the photometric redshift
relation having a tight correlation with spectroscopic redshift,
though about 0.03 below the one-to-one relation). It is clear,
however, that there remain a population of galaxies for which the CWW
templates do not well match the galaxy colors, leading to an over
estimate of the redshift of the galaxy.}
\label{fig:cww}
\end{figure}

\begin{figure}
%\plotone{figs/zzBC0.eps}
\plotone{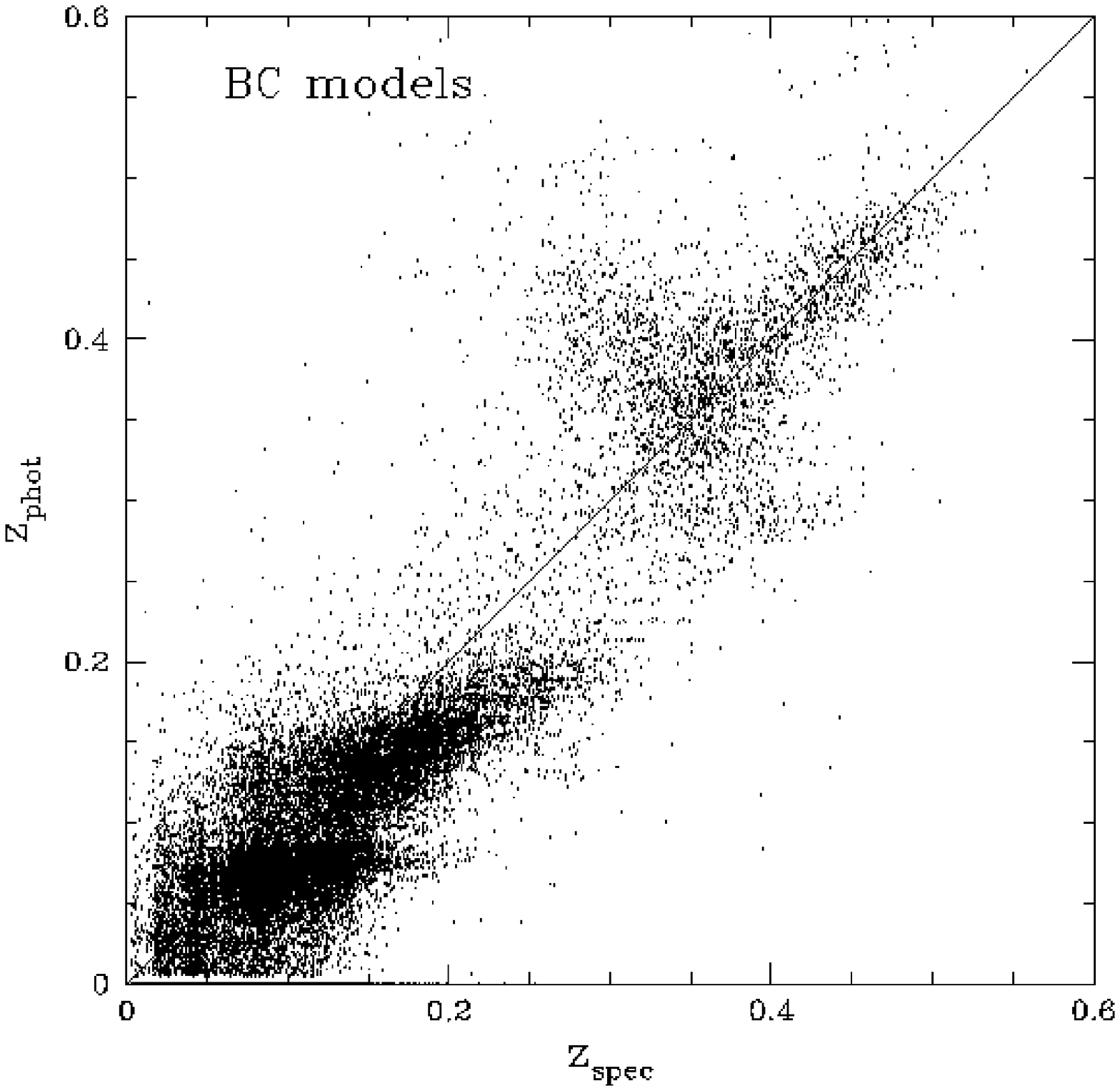}
\caption[]{Photometric redshift estimation using the Bruzual and
Charlot spectral energy distributions. The rms dispersion about this
relations is 0.051 in redshift. While this dispersion is within a
factor of 2 of that derived from adapting the templates it is also
clear that there remain systematic offsets within the photometric
redshift relation with the Bruzual and Charlot templates
under-predicting (on average) the true spectroscopic redshift.}
\label{fig:bc}
\end{figure}

The most frequently used set of spectral energy distributions (SEDs) used in
photometric redshift analyses are those from \citet[hereafter
CWW]{cww} (see also \citet{bolzonella00}). In Figures~\ref{fig:cww} and
~\ref{fig:bc} we demonstrate the results of the template fitting
technique using the CWW templates and a set of SEDs from the spectral
synthesis models of \citet{bc93}. The dispersion
about this relation is 0.062 and 0.051 for the CWW and BC templates
respectively.  While this is only a factor of two worse than that
achieved by the empirical methods there appear to be systematic
deviations within these photometric redshift relations. The CWW
templates produce a photometric redshift relation where the majority
of galaxies have a systematically lower redshift than that given by
the spectroscopic data (by approximately 0.03 in redshift) and there
exists a broad tail of galaxies for which the photometric redshifts
are systematically overestimated. For the BC templates the galaxy
redshifts tend to be systematically underestimated (with this effect
becoming more pronounced as a function of redshift out to redshifts
z=0.3).

\begin{figure}
%\plotone{figs/zzBPZ.eps}
\plotone{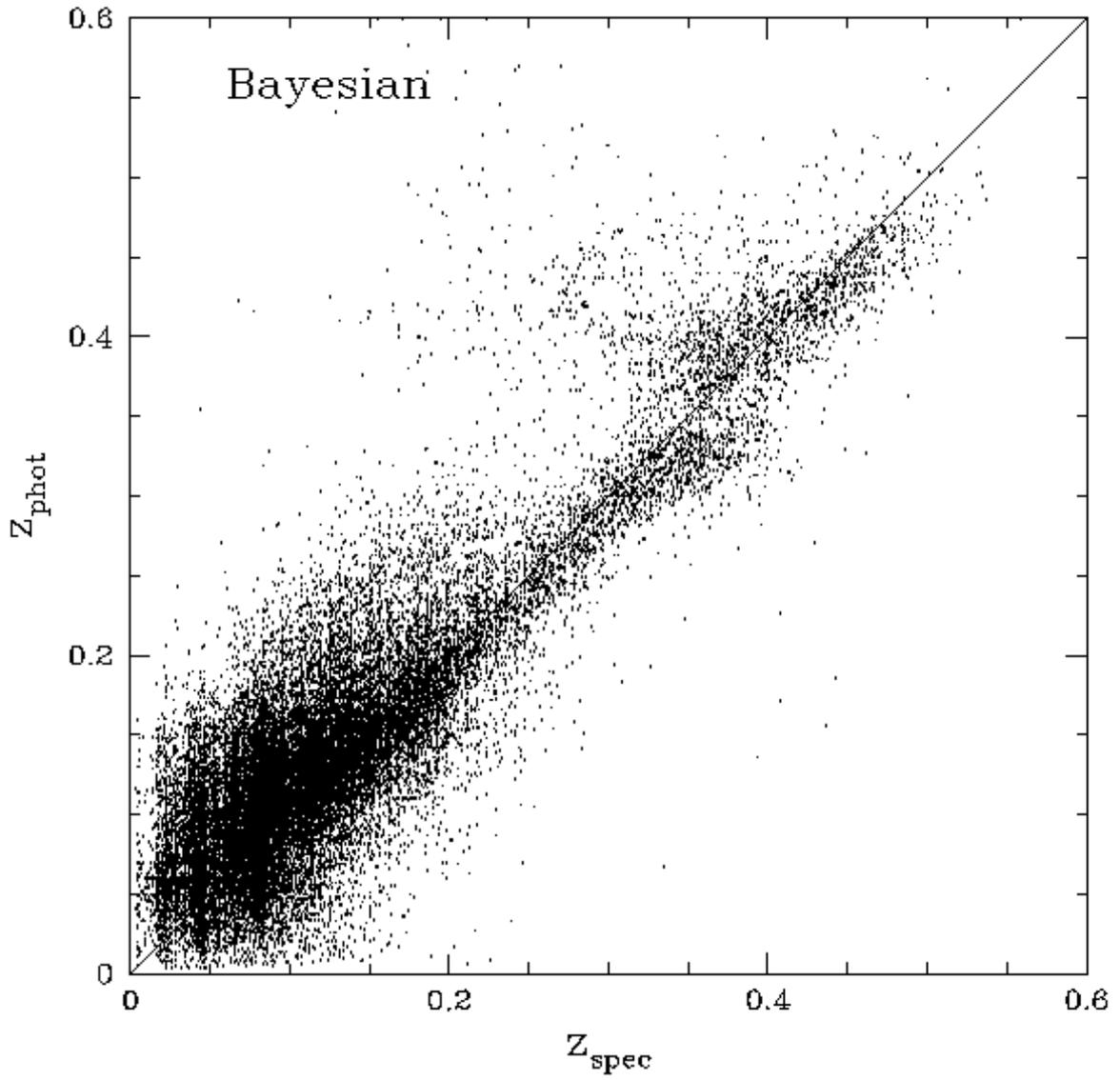}
\caption[]{Photometric redshift estimation using the Bayesian method.
The rms dispersion about this
relations is 0.042 in redshift. }
\label{fig:bpz}
\end{figure}

 An improvement over standard template methods, which rely uniquely on 
 the galaxy colors, is the introduction of magnitude priors within a 
 Bayesian framework (\citet{benitez00}).  The redshift distribution of the 
 main EDR sample is well fitted by the relationship $p(z)\propto  z^2 
 exp[-(z/z_m)^{1.5}]$ for $i \lesssim 18$, and a continuous prior can be 
 constructed by we measuring $z_m$ in 5 different magnitude bins and 
 interpolating. Since the EDR spectroscopic sample redshift distribution
 is 'contaminated' by LRGs at faint magnitudes and turns bimodal, we
 have assumed a flat redshift/magnitude prior for $i\gtrsim 18$. Using this 
 magnitude prior we run Bayesian estimation, with two further 
 refinements: a) setting the minimal photometric error in each band to 
 0.03, which mimics the intrinsic fluctuations in the colors of galaxies
 described by a same template and produces more realistic redshift 
 likelihoods and b) using linear interpolation between the main 
 CWW types to improve the color resolution. 
 Using this setup, the dispersion for the CWW templates {\it
 without using any prior} decreases from ~0.06 to ~0.05, with an offset of 
 0.0156; introducing the prior described above
 further decreases the dispersion to $\sigma_z=0.0415 $ 
 (see Figure~\ref{fig:bpz}) for the whole sample, but an offset of 0.0144 
 still remains. 

It is clear from these tests that while the template fitting methods
should be directly applicable to the SDSS EDR data there remain
significant systematics within either the templates or the photometric
calibrations (or both) that will add artifacts into any photometric
redshift relation. We must, therefore, recalibrate the template
spectra to minimize these systematic effects.

\section{Hybrid Photometric Redshift Techniques} 
\label{sec:hybrid}

Recently new hybrid techniques have been developed to calibrate
template spectral energy distributions (SEDs) \citet{csabai00,
budavari99, budavari00, budavari01} using a training set of
photometric data with spectroscopic redshifts. These combine the
advantages of the empirical methods and SED fitting by iteratively
improving the the agreement between the photometric measurements and
the spectral templates. The basic approach is to divide a set of
galaxies into a small number of spectral classes (using the standard
template based photometric redshifts) and then to adjust the template
SEDs to match the mean colors of the galaxies within these spectral
classes. By repeating this classification and repair procedure the
template spectra converge towards the observed colors. In this paper
we will not review the details of these techniques but direct the
reader to \citet{csabai00, budavari99, budavari00, budavari01} for a
full description of the algorithms. As we shall show in the following
sections the application of these techniques yields more reliable
photometric redshifts for the SDSS EDR catalog than the standard
template fitting.

\subsection{A Single Template: The Luminous Red Galaxy Sample} 
\label{sec:lrg}

In addition to providing a training set for redshift estimation within
the SDSS data the LRG sample is extremely useful in identifying
systematic uncertainties within the SDSS photometric system.  The LRG
galaxies have a strong continuum feature, namely the break at
around 4000\AA{}. Due to the depth of this feature, photometric
redshifts are easily estimated for these galaxies. In addition, due to
the high luminosity of these galaxies they can be observed,
spectroscopically over a larger redshift range than the main galaxy
sample. Systematics within the photometric data can, therefore, be
identified as this spectral feature passes through the filters as a
function of redshift.  In fact, we can simply use a single SED for the
LRG sample to test how we must optimize the template spectra to
accurately represent the observed colors.

\begin{figure}
%\plotone{figs/ellseds.eps}
\plotone{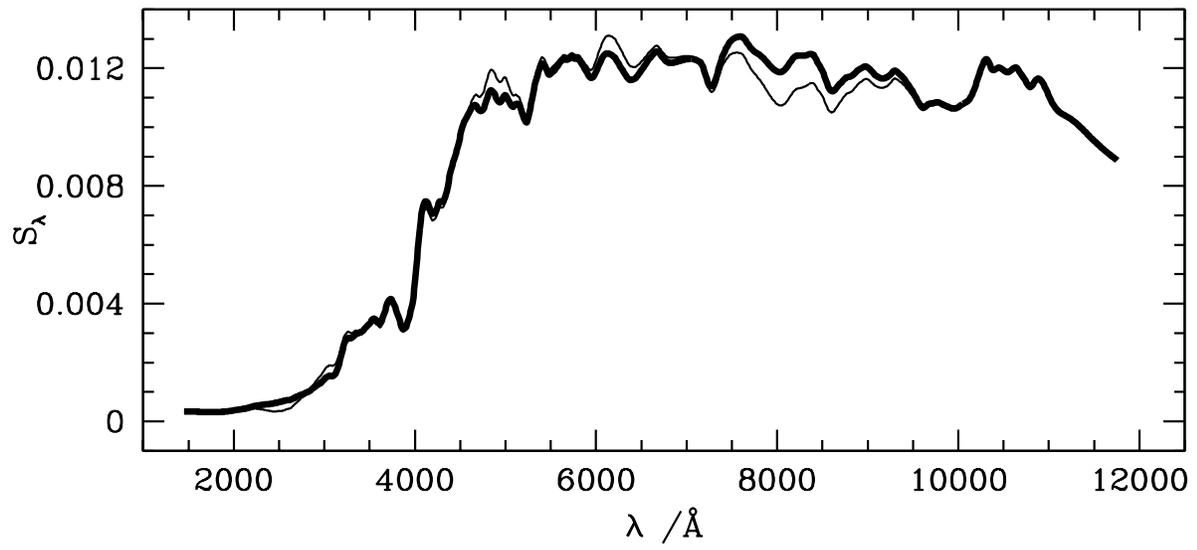}
\caption[]{The repaired (thick line) spectral template is redder than
the original elliptical galaxy template (thin line).}
\label{fig:ellseds}
\end{figure}

\begin{figure}
%\plotone{figs/color.eps}
\plotone{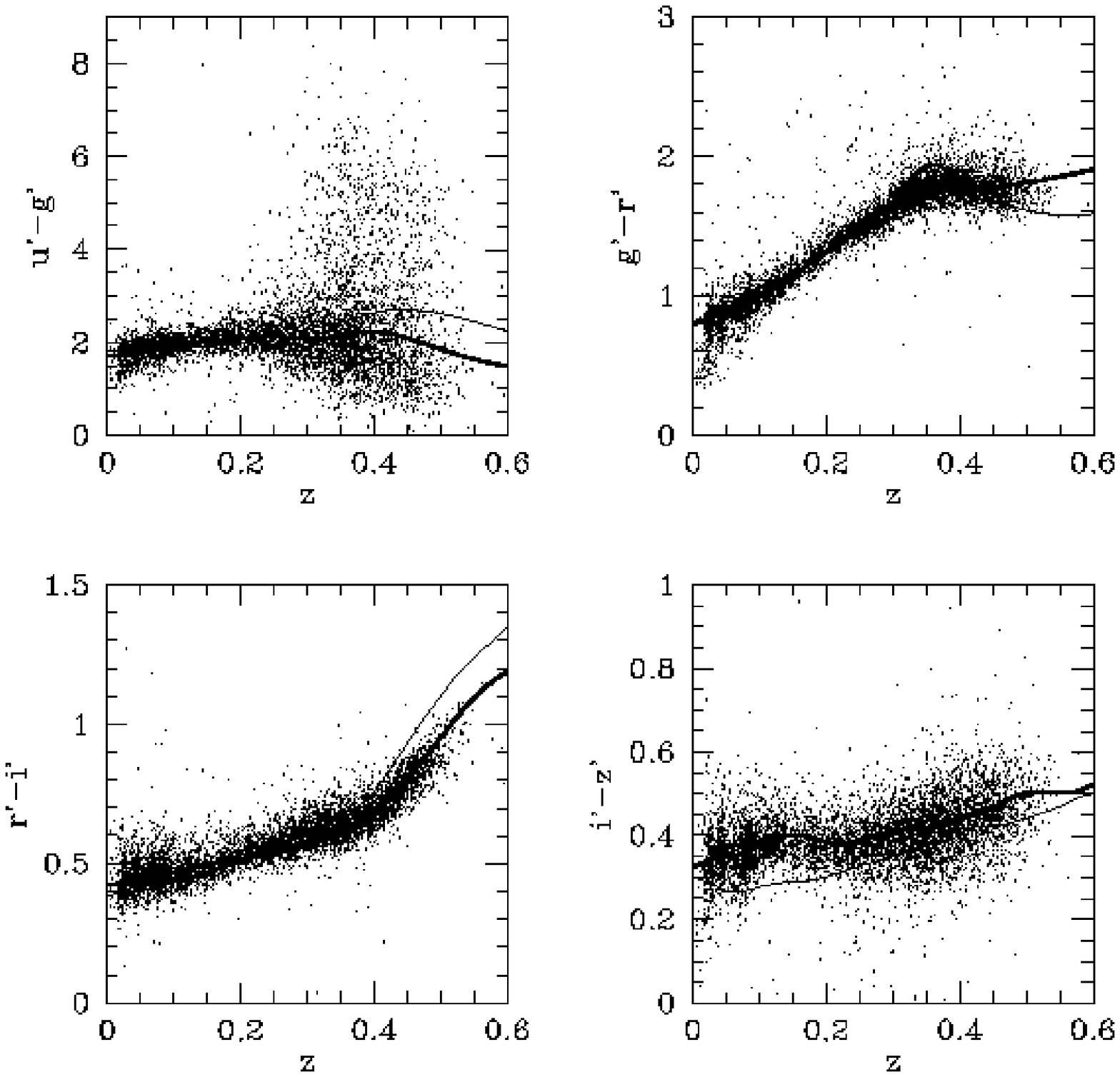}
\caption[]{ 
The four SDSS colors of $\sim$ 6000 red galaxies vs. the redshift. The
color trace of the repaired  spectral template (thick line) follows better the
data than the trace of the original CWW E0 template (thin line).}             
\label{fig:colortraces}
\end{figure}

For the 6698 LRG galaxies we start with an initial template spectrum
selected from the CWW elliptical spectrum and apply the training
techniques of Budavari et al (2000).  In Figure~\ref{fig:ellseds} we
show the original CWW elliptical spectrum together with
our reconstructed template. From these spectra we can
see that in oder to represent the colors of the LRGs we need a
template spectrum that is redder than the standard CWW elliptical.  To
demonstrate, how well these respective spectral templates cover the
photometric observations, we have plotted, in
Figure~\ref{fig:colortraces}, the colors of the EDR LRG galaxies
together with the traces of the original and repaired spectral
templates. The color-redshift relation for the repaired spectrum
clearly traces the locus of the LRG galaxy sample more accurately than
the original CWW SED. The most obvious improvement in the comparative
colors is found in the \ug{} and \iz{} colors.

\begin{figure}
%\plottwo{figs/zzLrgCww.eps}{figs/zzLrgRepMean10.eps}
\plottwo{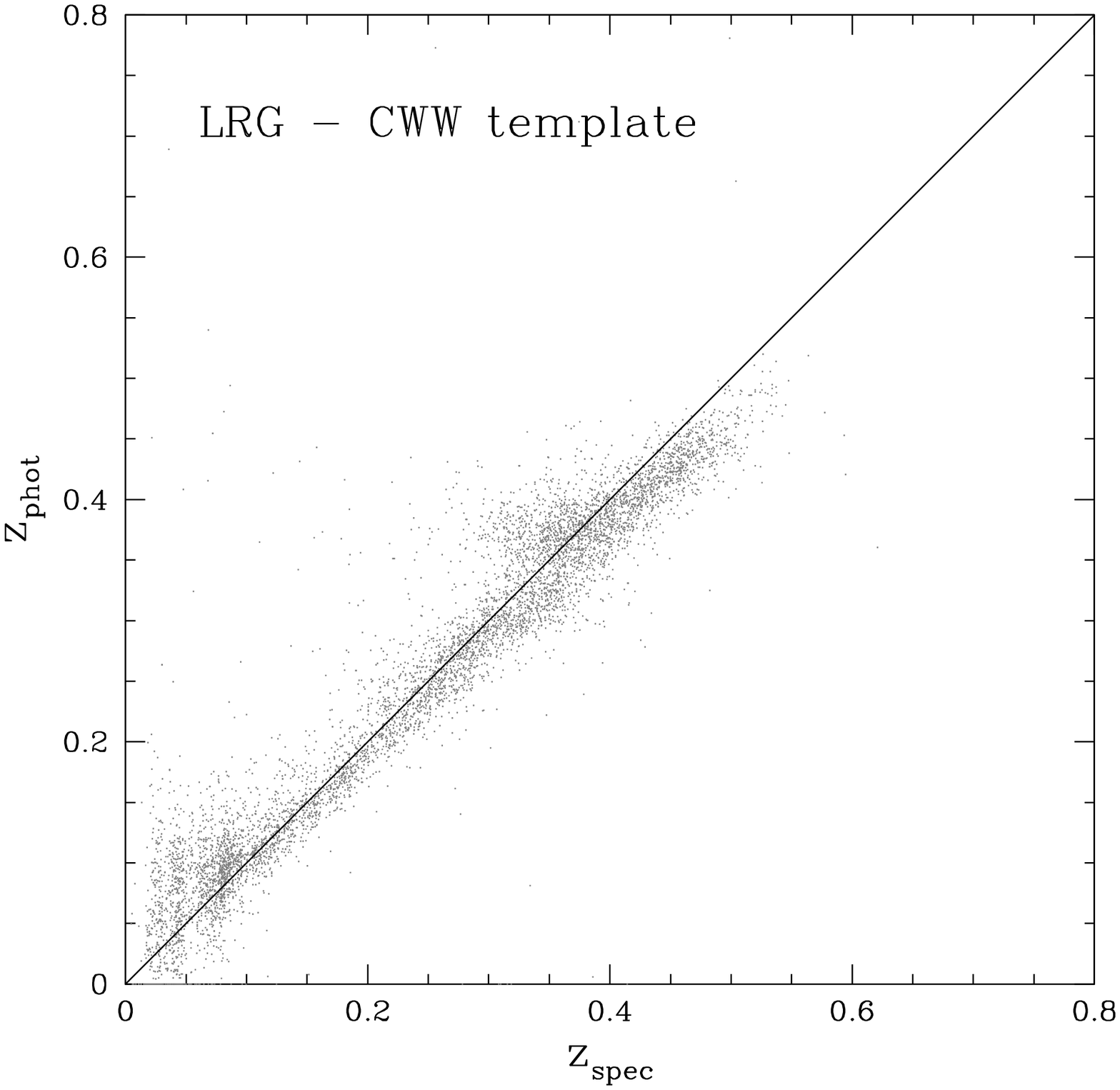}{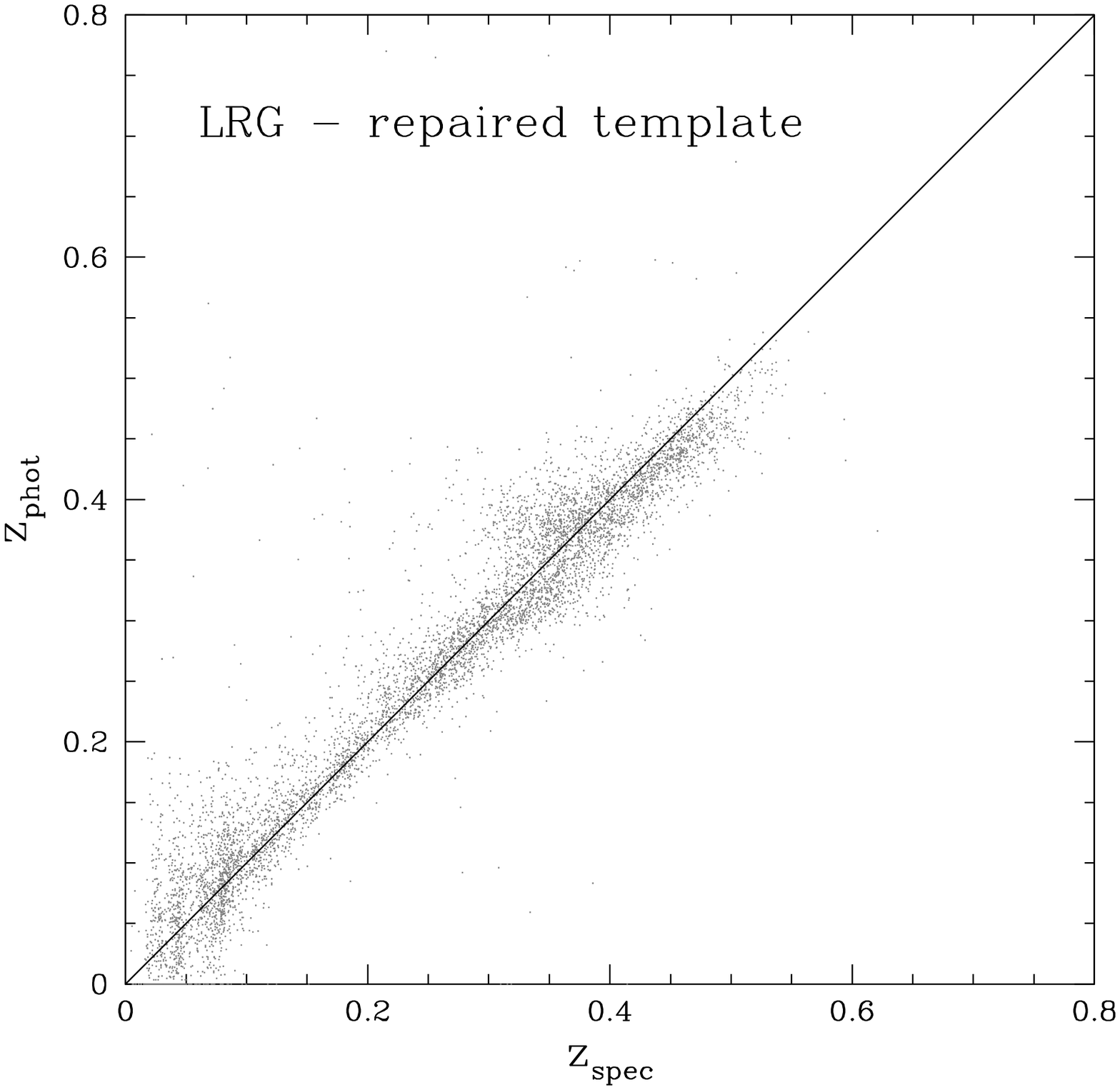}
\caption[]{Photometric vs. spectroscopic redshifts for the EDR LRG set.
On the left figure we used the original CWW spectral templates, while in
the figure on the left the templates were repaired. One can see, that
the redshift prediction improves, especially for higher redshifts.
}
\label{fig:templateLrg.figzz}
\end{figure}

Although the repair procedure does not optimize directly for
photometric redshifts, the improvement in the match between the
observed and predicted colors should lead to an improved photometric
redshift relation for the LRG sample.
Figure~\ref{fig:templateLrg.figzz} compares the performance of the
photometric redshift estimators utilizing the two original and
repaired template SEDs. The repair procedure decreases the overall
scatter in the redshift relation from $\sigma_z=0.031$ to
$\sigma_z=0.029$. The main improvement is, however, that
the systematic underestimation of the redshift, at redshifts $z>0.2$,
is reduced. There remains a feature in the redshift relation at $z
\approx 0.4$, an increase, by a factor of two, in the dispersion.
This arises due to the fact that there exists a degeneracy in the
$u'-g'$ vs $g'-r'$ colors within red galaxies at a redshift of
$z\sim 0.4$ (the color-color tracks loop on top of each other).  The
degeneracy is a result of the Balmer break shifting between the $g'$
and $r'$ filters making it difficult to estimate the exact redshift
(\citet{budavari01b}).  This problem cannot be removed by using better
template spectra.

\subsection{The Distribution of Galaxy Types: The Main galaxy Sample}
\label{sec:type}

The entire sample of the SDSS galaxies (including the LRGs) poses a
more difficult question due to the spectral composition of the data.
Spectral variations cannot be neglected and, in fact, one would like
to get a continuous parameterization of the spectral manifold. To
accomplish this we adopt a variant of the ASQ algorithm
(\citet{budavari01}). First we reconstruct a small number of discrete
SEDs using the techniques described previously and then we use an
interpolation scheme to provide a continuous distribution of spectral
types that evenly sample between the discrete spectra.

The training set consists of all galaxies with spectroscopic redshifts
and the 5 band SDSS photometry. The large number of galaxies is very
promising but the spectral resolution of the reconstructed templates
also depends on the redshift baseline of the input galaxy training
set. This redshift range is significantly smaller than, for example,
those derived from the Hubble Deep Field (\citet{hogg98,
budavari00}). Ideally, one would like to have a training set that
uniformly samples the color space to ensure that no extra weight is
assigned to any particular type of galaxy. The limited color range of
the galaxies with spectroscopic redshifts will, therefore, ultimately
limit the accuracy of our final redshift relations. 

\begin{figure}
%\plotone{figs/talkTest.eps}
\plotone{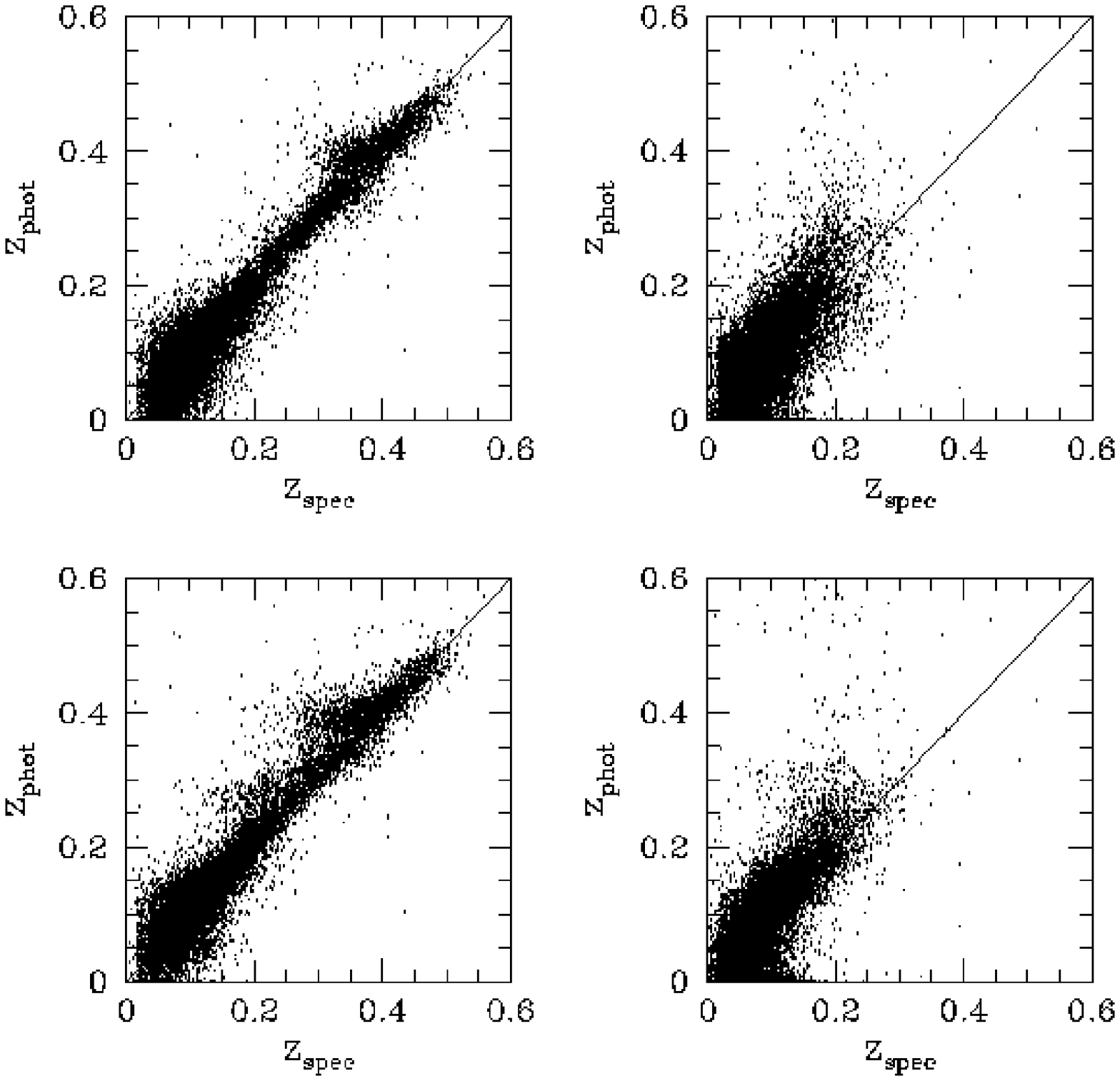}
\caption[]{Photometric redshifts of intrinsically red (panels on the left)
and blue (right panels) galaxies. The figures show the spectroscopic vs.\
photometric redshifts for the 4 discrete templates (top panels) and for the
continuous type (bottom panels) estimators. The type interpolation makes
things slightly worse for the red early-type galaxies because of the
type-redshift degeneracy, but the estimates for the late-type galaxies get
significantly better.  }
\label{fig:zzTemplate}
\end{figure}

The iterative ASQ method was applied to the initial set of four CWW
spectra. The spectral templates are found to converge rapidly, within
a few iterations. After 10 iterations, the repaired templates yield
photometric redshifts that are shown in the top panels of
Figure~\ref{fig:zzTemplate}. The left panel shows all galaxies
assigned to the reddest template and the galaxies assigned to the
remaining three templates are given on the right panel. The rms in the
red and blue sample are $\sigma_z = 0.028$ and $0.05$,
respectively. This plot should be compared with the redshift relations
derived from the standard CWW templates as shown in
Figure~\ref{fig:cww}. The training of these templates removes both the
systematics within the data and reduces the dispersion about the
photometric-redshift relation.

\begin{figure}
\epsscale{1}
%\plotone{figs/seds.eps}
\plotone{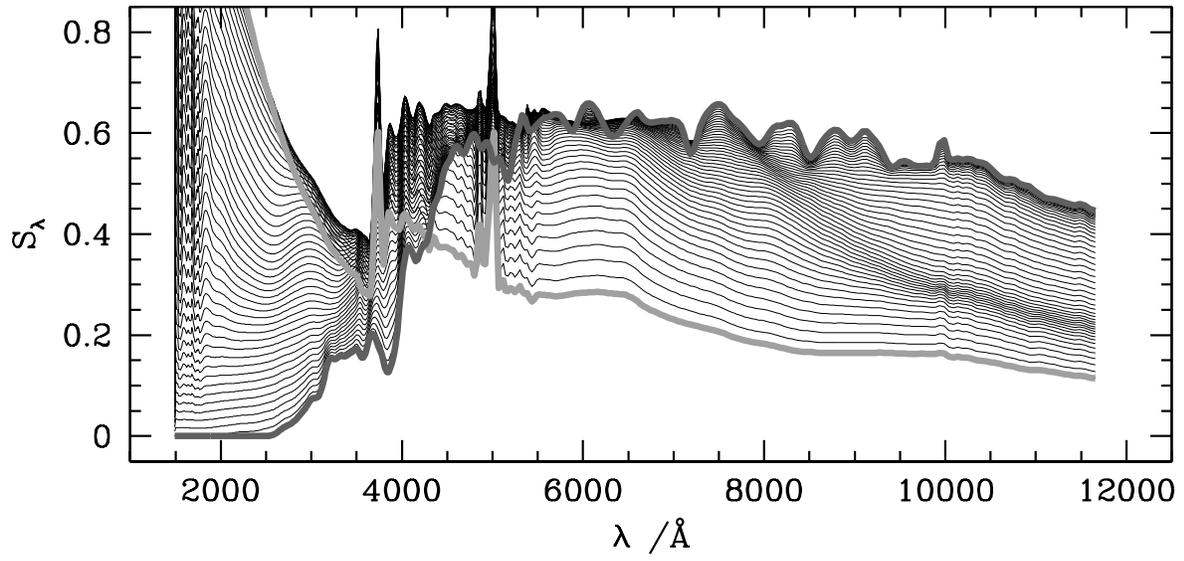}
\caption[]{Illustration of the 1D type manifold. A few SEDs are plotted here
for a equally spaced type parameter values. The reddest and bluest SEDs
are shown with the thick dark and light grey curves, respectively.}
\label{fig:seds}
\end{figure}

%%% TODO
%%%Analyzing the likelihood surfaces (see Figure~\ref{fig:chi1}) in the fits, 
%%%part of the increase in the dispersion for the late type galaxies comes from 
The large estimation error for the late type galaxies partly caused by 
the small number of discrete templates used in the redshift estimation. 
We can improve on our estimates if we derive
an interpolation scheme that provides a finer sampling of the
distribution of late type spectral templates. Figure~\ref{fig:seds}
illustrates the 1D continuous spectral manifold derived from the
discrete SEDs by plotting equally spaced (in type) interpolated
spectra using a simple spline interpolation. Based on the following
tests this simple interpolation scheme provides sufficient accuracy
for mapping the color distribution of late type galaxies. 

\begin{figure}
%\plotone{figs/typehist.eps}
\plotone{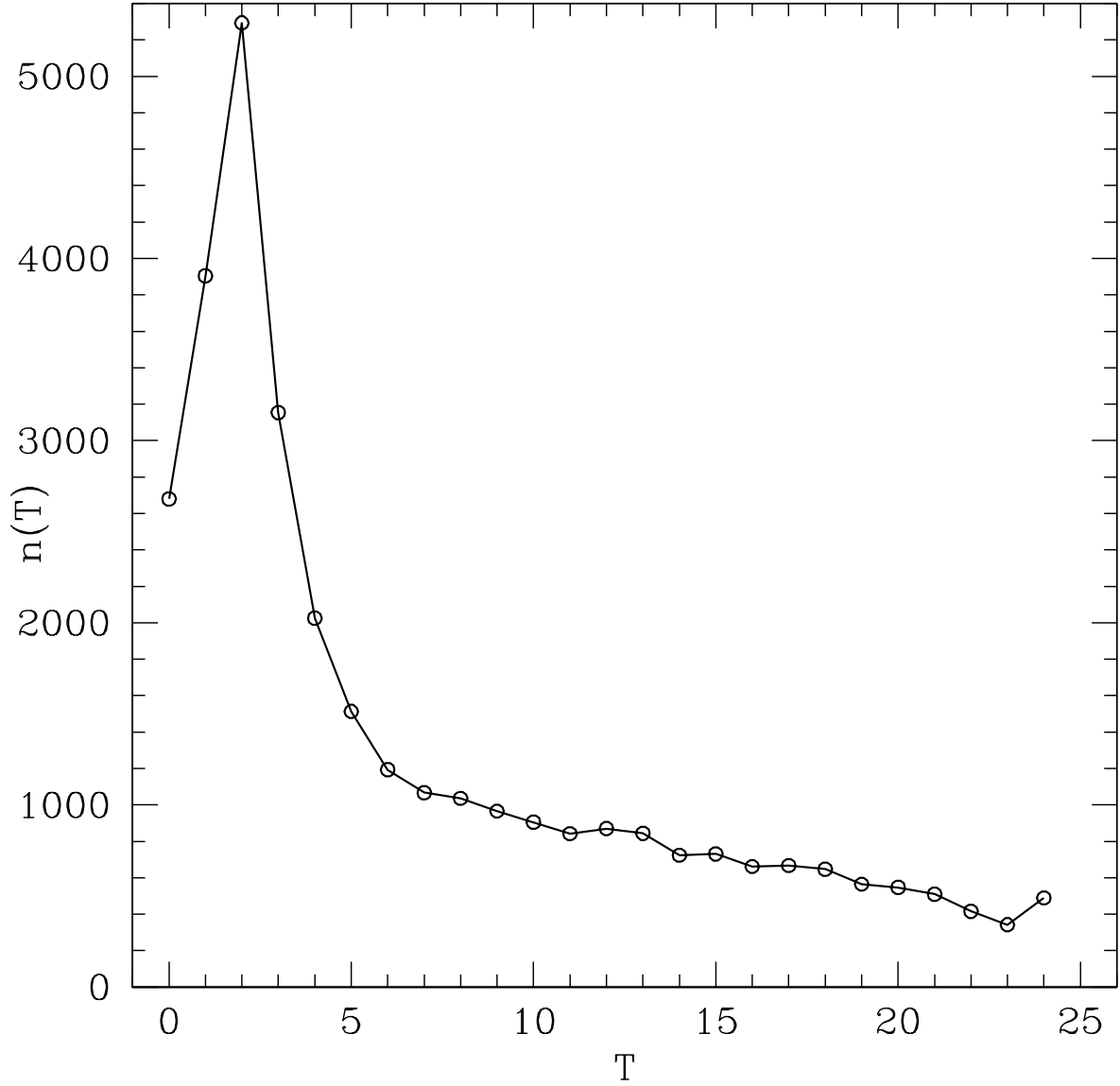}
\caption[]{The distribution of interpolated spectral templates that
fit the observed colors in the EDR main galaxy sample. The smooth
distribution shows that no particular spectral template is preferred
(i.e.\ the galaxies do not fall into a small number of spectral
types). This implies that the spline used to interpolated between the
trained spectral energy distributions accurately maps the distribution
of galaxy colors.}
\label{fig:type}
\end{figure}

The first test of the interpolation scheme was a simple sanity check
of the type histogram. If the interpolated spectra are not physical,
we expect to see humps at the basis templates (i.e.\ the colors of the
majority of galaxies will be better matched to the original templates
than the interpolated templates). For this test, we used the known
redshift of each galaxy in the training set and only fit the spectral
type (and apparent luminosity). In Figure~\ref{fig:type} we show this
interpolated type histogram. The smooth transition between
interpolated types shows no evidence for any discreteness in assigning
a spectral template to an individual galaxy. The second test of the
interpolation was to determine if the interpolated templates would
evolve if we applied the ASQ training algorithm.  Fixing the four basis
trained SEDs, we introduced three interpolated classes at the center of
the intervals between these spectral types. We find no significant
change in the spectral properties of these interpolated spectra as a
function of iteration of the training algorithm.

The redshift estimates based on the continuous 1D type parameter are
shown in the bottom panels of Figure~\ref{fig:zzTemplate} for both the
early- and late-type subsamples (left and right,
respectively). Compared to the top panels of the discrete version
(discussed previously), the new estimates seem to be superior for the
intrinsically blue subset and slightly worse of the early-types. 

For early-type galaxies it would be better to use the original discrete
template set to avoid the systematic overestimation around $z=0.2$ and
$z=0.3$. Since we want to have a simple estimation for the spectral
type, we would like to avoid to use a separate (discrete) template set
for early-type galaxies, so we use the above scheme keeping in mind the
systematic errors, and working on a better interpolated template set. 
Note, that SDSS will measure spectroscopic redshift
for most of the luminous early-type galaxies, so the number of objects
where this problem arises is somewhat smaller than in our test sample.
Though for the less luminous early type galaxies the above problem still exist. 

In
terms of rms values of the scatter this translates to an increase from
$\\sigma_z=0.028$ to $0.029$ for the red galaxies and a
decrease from $\sigma_z=0.05$ to $0.04$ for the blue
ones. To quote an rms for the entire training set would not be to
meaningful because it depends on the ratio of the number of early- and
late-type galaxies. For the main SDSS galaxy sample the scatter is
$\sigma_z=0.035$. We will use the above template fitting method with
repaired interpolated templates to create the EDR photometric
redshift catalog.

%{\tt  Do we need these in a table? I've put the numbers in the text...\\
%--- RMS ERROR RESULTS --- \\
%  4SED  ALL  0.0473 32592 0.0383 32592 0.0390 98.5\% \\
%  4SED  RED  0.0332 18569 0.0285 18569 0.0283 97.6\% \\
%  4SED  BLU  0.0611 14023 0.0509 14023 0.0500 98.8\% \\
%  TYPE  ALL  0.0451 32592 0.0359 32592 0.0352 97.7\% \\
%  TYPE  RED  0.0363 18569 0.0302 18569 0.0291 96.9\% \\
%  TYPE  BLU  0.0547 14023 0.0433 14023 0.0419 98.1\% \\
%}

\section{Comparisons with Independent Redshift Samples}
\label{sec:blind}

\subsection{The 2dF and CNOC2 Redshift Samples}

\begin{figure}
%\plottwo{figs/2dfzz.eps}{figs/cnoc2.eps}%deepzz.eps}
\plottwo{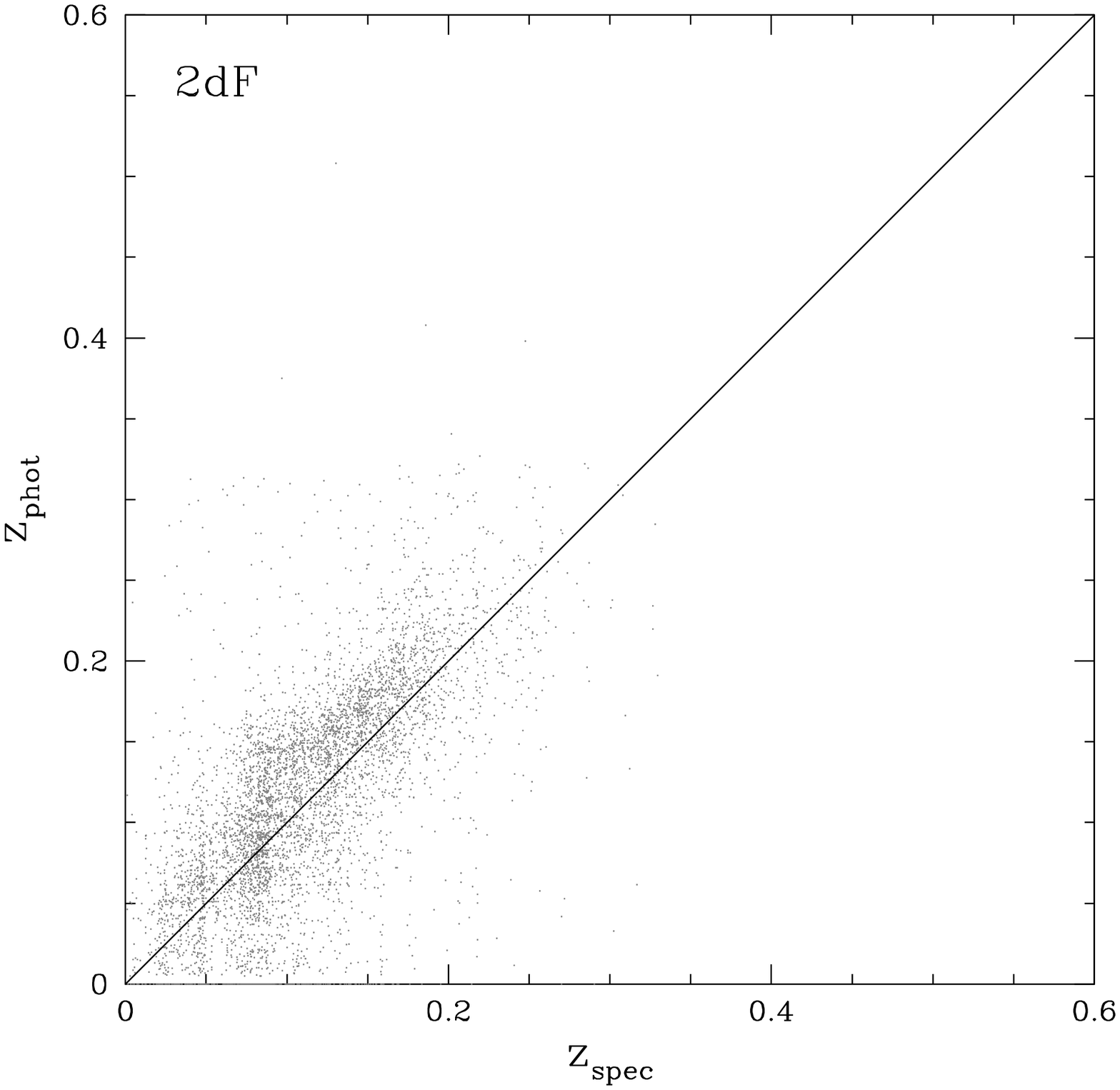}{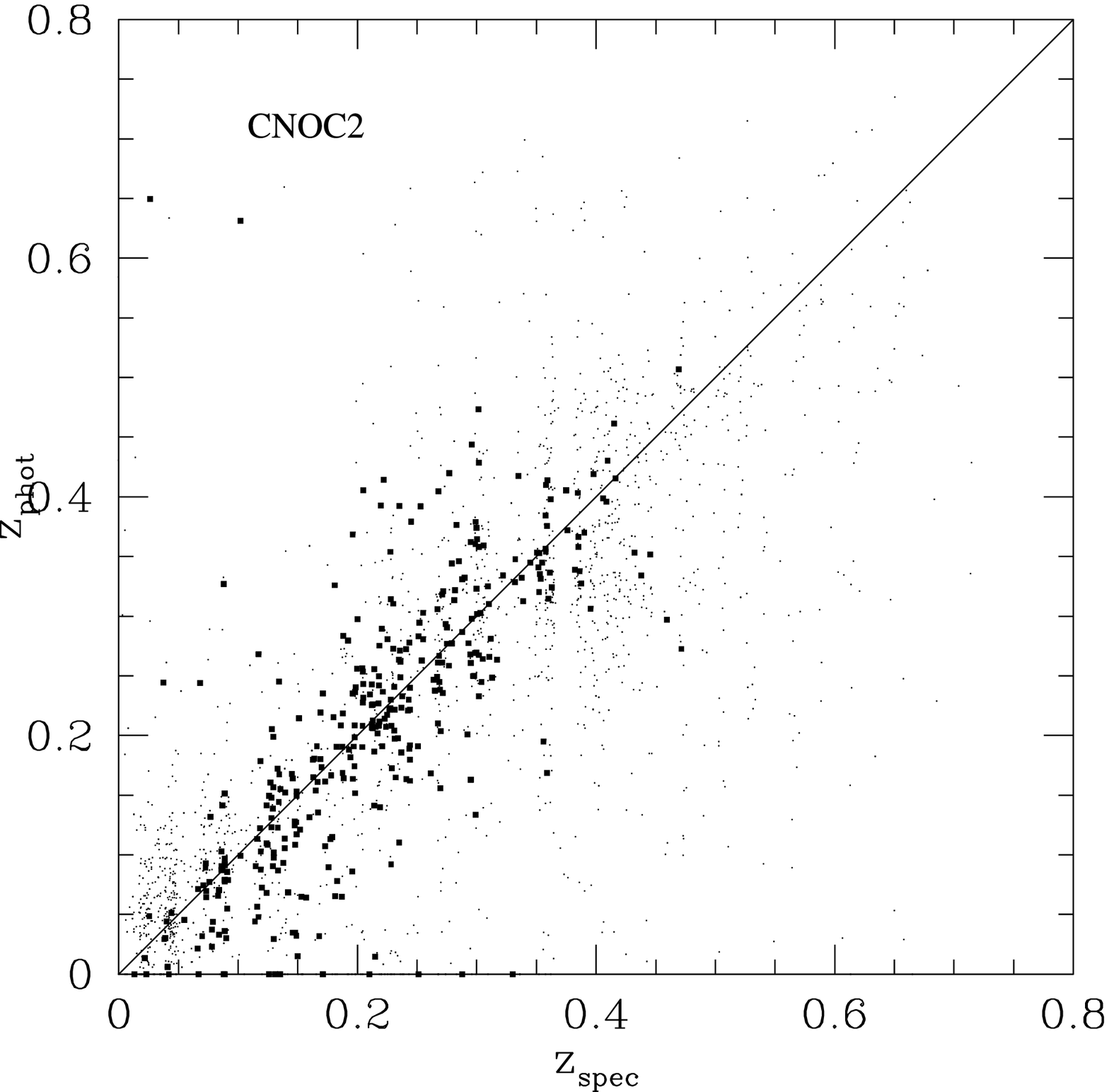}
\caption[]{ 
Checking the extrapolation capabilities of the photometric redshift
estimator: the predicted vs. the spectroscopic redshift.  Left: 2dF
set. Right: The CNOC2 set; since most of these objects are too faint, we
show with larger symbols the objects with reasonable
SDSS photometry ($17.8<r<19.5$). Note the different redshift range. 
}             
\label{fig:blindzz}
\end{figure}

In the above sections we have used data from the same subsets for
training and testing. We now perform a blind test using the
independent data sets. Details of the 2dF and CNOC2 data sets are
given in Section~\ref{sec:data}. Figure~\ref{fig:blindzz}a compares
the spectroscopic and photometric redshifts for the 2dF spectroscopic
sample. The dispersion within the photometric-redshift relation for
these data is, $\sigma_z = 0.043$. This compares to the
dispersion in the relation for the full SDSS sample of $\sigma_z
 = 0.035$. The increase in the dispersion arises from two
effects. The $r'$ band magnitudes of the 2dF data are intrinsically
fainter than the SDSS spectroscopic sample (by approximately 0.2
magnitudes) and the 2dF data are selected based on their $B_j$
photographic magnitudes which will provide an intrinsically bluer
galaxy sample than the $r'$ selected SDSS data. As the dispersion in
the redshift relation increases with limiting magnitude and for blue
galaxies the difference in the observed photometric redshift relation
is not surprising.

\begin{figure}
%\plotone{figs/rms.ps}
\plotone{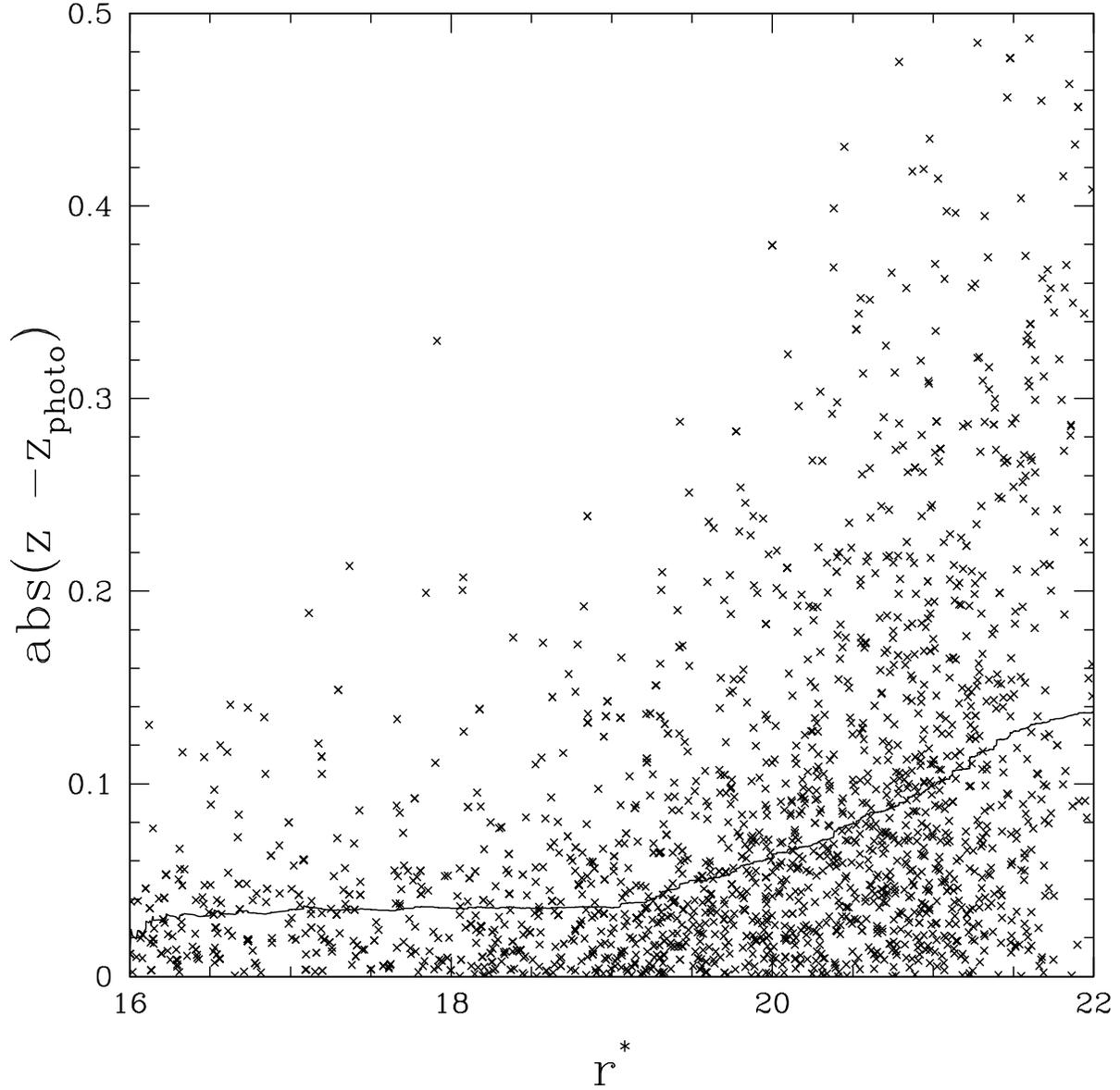}
\caption[]{ The cumulative rms of the SDSS photometric redshift as a
function of limiting magnitude. The points represent the absolute
deviation between the spectroscopic and photometric redshifts for the
CNOC2 sample of galaxies. The solid line is the cumulative rms of the
sample as a function of the $r'$ magnitude. At a limiting magnitude
of $r'<21$ the rms error on the photometric redshift rises to 0.1.  }
\label{fig:rms}
\end{figure}

To determine how well the templates extrapolate to higher redshift
data we apply the photometric redshifts to the CNOC2 data set (with a
redshift range $0<z<0.7$ and a magnitude limit of $r'<21.0$) As we
can see in Figure~\ref{fig:blindzz} the dispersion in the relation
increases for the fainter magnitude sample due to the increase in
photometric error. The average estimation error for the whole set is
$\sigma_z = 0.084$. If we consider only those galaxies with
$ 17.8<r'<19.5 $, the uncertainty in the redshift estimates decreases
to $\sigma_z = 0.061$. In Figure~\ref{fig:rms} we show the
absolute deviation between the photometric and spectroscopic redshifts
for the CNOC2 galaxy sample as a function of $r'$. The cumulative rms
of these data (as a function of $r'$) is shown by the solid line. For
$r' < 21$ the rms uncertainty about this relation is 0.1 in redshift.

\section{The Early Data Release Photometric Redshift Catalog} 
\label{sec:cat}

\subsection{Selecting Galaxies From the EDR Database}

The goal of our analysis has been to obtain photometric redshifts for
all SDSS galaxies in the Early Data Release. We have, therefore,
created the first EDR photo-z catalog {\em (version 1.0)} which has
now been included in the publicly available EDR database at
\myurl{http://skyserver.sdss.org/}.

We used the template fitting method with repaired interpolated templates
\ref{sec:hybrid} to estimate photometric redshifts in the above public catalog.
Though the empirical methods (see \ref{sec:emp}) give smaller estimation
error, we have chosen to use the template fitting method since it estimates
not just redshift, but spectral type and restframe magnitude, too. Also
we hope, that with the accumulation of more precisely calibrated data in 
further SDSS releases, the disadvantage of this method decreases.

The photometric redshift table (see Table~\ref{tab:param.tbl} for the list of
parameters) in the database has more than 6 million
entries, one for every galaxy in the EDR. Each entry contains the unique
object ID (\p{objID}, for quick cross-matching), the most likely redshift
(\p{z}) and type (\p{t}). The uncertainties of redshift and type
calculated from the 68\% confidence regions of the fit assuming Gaussian
errors. Note that the true error distribution for higher redshift object
is not known, and probably not Gaussian. 
The elements of the covariance matrix are stored in the database
and represented by \p{c\_zz}, \p{c\_tt}, \p{c\_tz}. The errors in columns
\p{zErr} and \p{tErr} are simply taken from the diagonal elements of the
covariance matrix.
The $\chi^2$ value of the fit (\p{chiSq}) measures the
absolute `goodness' of the fit. The catalog contains a preliminary quality
flag (\p{quality}), which scales between zero and five where the larger
the number the more confident the photometric redshift. This flag is
assigned to objects in the process of fitting the confidence region and
seems to correlate with the rms of the photometric and spectroscopic
redshifts. In the current version this correlation is quite weak, we would 
like to improve the calculation of this flag in the next version. 

In addition to the redshift estimates physical parameters derived from the
estimated redshift are also stored within the database. These include the
distance modulus (\p{dmod}) for the standard $\Lambda$CDM cosmology
($\Omega_{\rm{M}}=0.3$, $\Omega_{\Lambda}=0.7$, $h^{-1}$ units), restframe
colors (\p{rest\_ug}, \p{rest\_gr}, \p{rest\_ri}, \p{rest\_iz}) and
K-corrections (\p{kcorr\_u}, \p{kcorr\_g}, \p{kcorr\_r}, \p{kcorr\_i},
\p{kcorr\_z}) derived directly from the templates and the restframe
absolute magnitudes (\p{absMag\_u}, \p{absMag\_g}, \p{absMag\_r},
\p{absMag\_i}, \p{absMag\_z}) as computed from the distance modulus and
K-correction, $$ M = m - {\rm D\!M}(z) - {\rm{}K}(t,z) . $$

Access to these parameters is straightforward through the Structured
Query Language (a.k.a. SQL). A sample query to extract the objId and
photometric redshift of 5 galaxies in the redshift range of
$0.2<z<0.3$ would look like this:

\begin{center}
\begin{minipage}[0]{5cm}
\begin{verbatim} 
select top 5 objId, z  
   from PhotoZ 
   where z>0.2 and z<0.3
\end{verbatim}
\end{minipage}
\end{center}

All parameters stored within the SDSS database (including the derived
parameters) can be searched upon.

\subsection{Caveats and Limitations of the Current Photometric Redshifts} 
\label{sec:caveats}

While, as the comparisons between the photometric and spectroscopic
redshift show, the current implementation of SDSS photometric
redshifts provide an accurate estimate of the redshifts there are a
number of limitations and caveats pertaining to the EDR data. We
describe here the results of a series of tests of the quality of the
SDSS photometry and how these issues affect the accuracy and possible
uses of the photometric redshifts in the EDR catalog. We advise any
potential user of the current photometric redshift implementation to
be aware of these caveats prior to undertaking any statistical
analysis.

Even though the photometric calibration of the SDSS survey has been
shown to be be accurate to a few percent for the SDSS standard stars,
galaxy colors appear to have a slight offset from SED based estimated
values (\citet{eisenstein01}). As part of this analysis of the SDSS EDR
data we compare measured colors not only to the spectrophotometrically
calibrated SEDs (e.g.\ CWW) but we have also carried out experiments
where small offsets were applied before the refining the template
spectra. In this way we can identify systematic photometric offsets
from the mean deviation of the colors from the SEDs.  The $g'$ band
offset we found is in the same sense as that given in
\citet{eisenstein01} but with a smaller amplitude of $\Delta g' \sim
0.05$. All galaxies within the SDSS catalog had this $g'$ offset
applied prior to calculation of the photometric redshifts.

Our SED reconstruction algorithm ideally requires a training set with
reasonably uniform redshift distribution over a large baseline. The
SDSS spectroscopic survey delivers excellent quality data for this
kind of analyses. However, the main galaxy sample has a median
redshift of approximately 0.1 which does not enable the use
photometric data from different bands to constrain the SEDs at all
wavelengths. In principle, if there exist photometric zeropoint
uncertainties within the data, the reconstruction could introduce
artificial continuum spectral features in the templates that would
make the extrapolation to higher redshifts impossible (in a similar
sense to the limitations of the empirical techniques). The repaired
spectral energy distributions show no obvious trace of such
features. 

\begin{figure}
%\plotone{figs/zhist.eps}%caveathist.eps}
\plotone{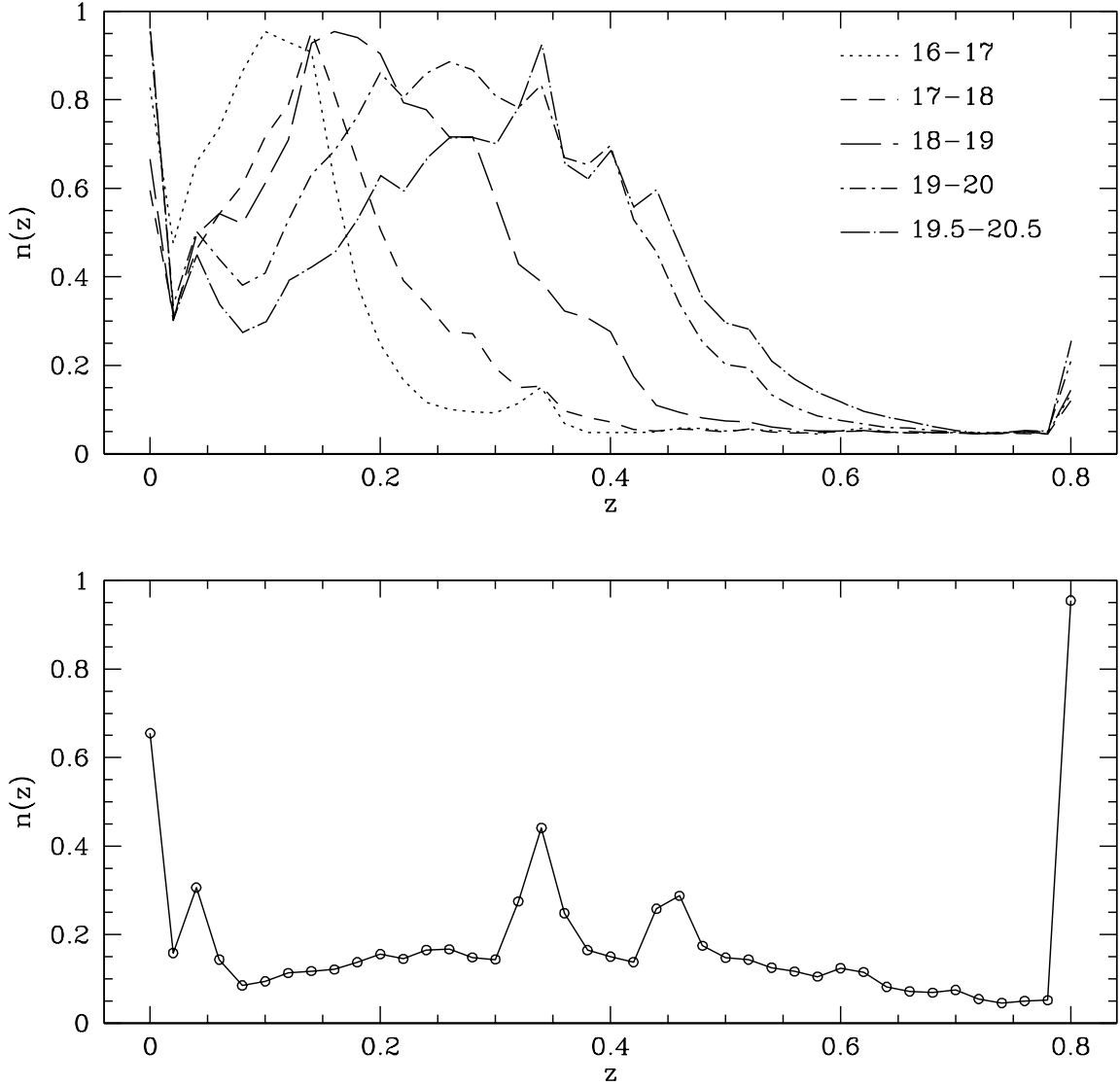}
\caption[]{These redshift distributions are plotted here to show the
trend with the apparent $r'$ band magnitude. As expected, the
histograms in the upper figure are shifted to right as we go with
$r'$ magnitude binss form $16<r'<17$ to $19.5<r'<20.5$.  The
histogram in the lower figure is built using all galaxies in the EDR
catalog, and has artifacts. We, therefore, advise caution when using
the current EDR photometric redshift catalog for galaxies with
$r'>21$.}
\label{fig:caveathist}
\end{figure}

Finally, we consider how the increasing photometric uncertainty at
fainter magnitudes affect the redshift histograms. In
Figure~\ref{fig:caveathist} we show the redshift distributions in
different $r'$ magnitude bins 16--17, 17--18, 18--19, 19--20 and
20--21. The histograms built in different magnitude bins peak around
values consistent with published redshifts surveys and that move
toward higher values as a function of the magnitude.  Beyond a
magnitude limit of $r'>21$ artifacts are seen within the redshift
histograms due to the large photometric errors. We, therefore, advise
caution when using the current EDR photometric redshift catalog for
galaxies with $r'>21$. Also one should take into consideration the fact
that for some objects the photometric redshift would be negative because 
the estimation is based on photometric data with errors, but the
algorithm allows only positive redshift values, so all negative
redshifts pile up at $z=0$.

\section{Conclusions} 
\label{sec:concl}

We present the first application of photometric redshifts to the SDSS
EDR data. From a comparison of the photometric and spectroscopic
redshifts we find that the rms error within the redshift relation is
$0.035$ for $r'<18$ rising to 0.1 at $r'<21$. For magnitude
intervals $r'<21$ the photometric redshift relation and redshift
histogram are well matched to existing redshift surveys (with
comparable median redshifts and dispersions). Implementing these
redshift estimates in the SDSS EDR database, together with derived
quantities such as the absolute magnitudes, k-corrections and
restframe colors, we provide a simple interface to one of the largest
publicly accessible catalogs of photometric redshifts available to the
astronomical community. We conclude by providing a description of the
limitations and caveats present within the current photometric
redshift implementation. We caution all users to be aware of these
limitations before applying the EDR photometric redshifts in any
statistical analyses.

\acknowledgments

Funding for the creation and distribution of the SDSS Archive has been provided
by the Alfred P. Sloan Foundation, the Participating Institutions, the National
Aeronautics and Space Administration, the National Science Foundation, the U.S.
Department of Energy, the Japanese Monbukagakusho, and the Max Planck
Society. The SDSS Web site is http://www.sdss.org/. 
The SDSS is managed by the Astrophysical Research Consortium (ARC) for the
Participating Institutions. The Participating Institutions are The University of
Chicago, Fermilab, the Institute for Advanced Study, the Japan Participation
Group, The Johns Hopkins University, Los Alamos National Laboratory, the
Max-Planck-Institute for Astronomy (MPIA), the Max-Planck-Institute for
Astrophysics (MPA), New Mexico State University, Princeton University, the
United States Naval Observatory, and the University of Washington.

I.C. and T.B.
acknowledge partial support from the MTA-NSF grant no.\ 124 and the
Hungarian National Scientific Research Foundation (OTKA) grant no.\
T030836.  A.S.  acknowledges support from NSF (AST9802980) and a NASA
LTSA (NAG53503). A.J.C.  acknowledges partial support from NSF grants
AST0096060 and AST9984924 and an NASA LTSA grant NAG5 8546.

\newpage

%%
%% Figures
%%

%\begin{figure}
%\epsscale{0.5}
%%\centerline{\plotone{figs/mapedit2.eps}}%figure.cov3.eps}}
%\centerline{\plotone{Csabai.fig10.eps}}
%\caption[]{ Template fitting techniques compare observed colors with
%those measured from spectral energy distributions. Typically the
%parameter space is searched using high resolution redshift grids and a
%handful of templates. In this figure we plot the probability map of a
%galaxy using high resolution grids in redshift (x-axis) {\em and}
%spectral type (y-axis). The gray-scale image illustrates the
%likelihood curve over the $(z,t)$ plain.  The most important thing to
%note is the fact that based on the photometric measurements of this
%particular object there are not a large number of local extrema. The
%likelihood function is well described by a global optimum. However, if
%one uses the standard approach, fitting say 5 discrete templates, then
%one would end up with a few local maxima of the probability as
%function of redshift and type, $p(z,t)$. In fact none of these maxima
%would be the optimal solution that is suggested by the photometry. The
%only solution to this problem is to use a continuous and finely
%sampled distribution of templates (i.e.\ a continuous manifold).}
%\label{fig:chi1}
%\end{figure}

\newpage
%% tables
%%

%~budavari/py/rms.py  edrSpec.photoz.poly2 
%0.0318 32592 0.0277 32592   0.0273 98.0%
%~budavari/py/rms.py  edrSpec.photoz.nnn   
%0.0365 32592 0.0321 32592   0.0327 98.5%
% ~budavari/py/rms.py edrSpec.types.photoz
%0.0254 32592 0.0224 32592   0.0226 98.4%
\begin{deluxetable}{lccccc}
 \tablecolumns{5}
 \tablewidth{0pc}
 \tablecaption{\sc Errors on Photometric Redshifts }
 \tablehead{
  \colhead{Estimation Method} &
  \colhead{rms} &
  \colhead{log} &
  \colhead{iterated} &
  \colhead{non-outliers} &
 }
 \startdata
   Polynomial        & 0.0318 & 0.0277 & 0.0273 & 98.0\% \\
   Nearest neighbor  & 0.0365 & 0.0321 & 0.0327 & 98.5\% \\
   Kd-tree           & 0.0254 & 0.0224 & 0.0226 & 98.4\% \\
   CWW               & 0.0666 & 0.0598 & 0.0621 & 99.1\% \\
   Bruzual-Charlot   & 0.0552 & 0.0501 & 0.0509 & 99.2\% \\
   Bayesian          & 0.0476 & 0.0415 & 0.0422 & 98.4\% \\		
   CWW LRG           & 0.0473 & 0.0332 & 0.0306 & 97.1\% \\
   Repaired LRG      & 0.0476 & 0.0319 & 0.0289 & 96.5\% \\
   Interpolated      & 0.0451 & 0.0359 & 0.0352 & 97.7\% \\
   2dF               & 0.0528 & 0.0455 & 0.0433 & 97.1\% \\
   CNOC2             & 0.1358 & 0.0989 & 0.0842 & 93.0\% \\
   CNOC2 $17.8<r<19.5$           & 0.0801 & 0.0614 &  0.0614 & 97.1\% \\
 \enddata
\label{tab:rms.tbl}

\tablecomments{ We list 3 different estimated rms values in the Table.  The
first is the usual standard deviation $\sigma_{\rm rms}$ computed for
all galaxies as defined by $\sigma_{\rm rms}^2 = \langle \Delta z^2
\rangle$, where $\Delta z=z_{\rm spec}-z_{\rm phot}$.  The standard
deviation is very sensitive to outliers, it is a common trick to assign
less weight to them by defining another quantity that measures the scatter
in a more reliable way: $\sigma_{\log}^2 = \left \langle A^2 \log
\left(1+\Delta z^2 / A^2 \right) \right \rangle$ where $A$ is a large
number compared to $\Delta z$. We use $A^2=20\times\Delta z_{\rm med}^2$,
where $\Delta z_{\rm med}$ is the median. Without outliers $\sigma_{\rm
rms}$ and $\sigma_{\rm log}$ were basically same, because $\epsilon
\approx \log(1+\epsilon)$ for small $\epsilon$ values, but large outliers
only affect the standard deviation drammatically. Another way of
suppressing the effect of outliers is excluding them. The last rms
 column ($\sigma_z$; we use this values in the text)
lists the standard deviation for galaxies that are within the $3\sigma$
limits of the distribution, which often has a value similar to
$\sigma_{\rm log}$. The very last column of the table shows the fraction
of galaxies included in the $3\sigma$ limit.}
\end{deluxetable}  

\begin{deluxetable}{llllll}
\tablecolumns{5}
\tablewidth{0pc}
\tablecaption{\sc Photometric Redshift Parameters }
\tablehead{
\colhead{name} &
\colhead{type} &
\colhead{length} &
\colhead{unit} &
\colhead{description} &
}
\startdata
$pId$	& int	&  4	&  -	&    unique Id for photoz version \\
$rank$	& int	&  4	&  -	&    the rank of the photoz determination, default is 0 \\
$version$	& varchar	&  6	&  -	&    the version of photoz code \\
$class$	& int	&  4	&  -	&    char describing the object type (galaxy:1, QSO:tbd, ...) \\
$objID$	& bigint	&  8	&  -	&    unique ID pointing to PhotoObj table \\
$chiSq$	& real	&  4	&  -	&    the chi-square value for the fit \\
$z$	& real	&  4	&  -	&    photometric redshift \\
$zErr$	& real	&  4	&  -	&    the marginalized error of the photometric redshift \\
$t$	& real	&  4	&  -	&    photometric SED type between 0 and 1 \\
$tErr$	& real	&  4	&  -	&    the marginalized error of the photometric type \\
$c_{tt}$	& real	&  4	&  -	&    tt element of covariance matrix \\
$c_{tz}$	& real	&  4	&  -	&    tz element of covariance matrix \\
$c_{zz}$	& real	&  4	&  -	&    zz element of covariance matrix \\
$fitRadius$	& int	&  4	&  pixels	&    radius of area used for covariance fit \\
$fitThreshold$	& real	&  4	&  -	&    probability threshold for fitting, peak normalized to 1 \\
$quality$	& int	&  4	&  -	&    integer describing the quality (best:5, lowest 0) \\
$dmod$	& real	&  4	&  magnitudes	&    distance modulus for Omega=0.3, Lambda=0.7 cosmology \\
$rest_{ug}$	& real	&  4	&  magnitudes	&    rest frame u-g color \\
$rest_{gr}$	& real	&  4	&  magnitudes	&    rest frame g-r color \\
$rest_{ri}$	& real	&  4	&  magnitudes	&    rest frame r-i color \\
$rest_{iz}$	& real	&  4	&  magnitudes	&    rest frame i-z color \\
$kcorr_u$	& real	&  4	&  magnitudes	&    k correction \\
$kcorr_g$	& real	&  4	&  magnitudes	&    k correction \\
$kcorr_r$	& real	&  4	&  magnitudes	&    k correction \\
$kcorr_i$	& real	&  4	&  magnitudes	&    k correction \\
$kcorr_z$	& real	&  4	&  magnitudes	&    k correction \\
$absMag_u$	& real	&  4	&  magnitudes	&    rest frame u' abs magnitude \\
$absMag_g$	& real	&  4	&  magnitudes	&    rest frame g' abs magnitude \\
$absMag_r$	& real	&  4	&  magnitudes	&    rest frame r' abs magnitude \\
$absMag_i$	& real	&  4	&  magnitudes	&    rest frame i' abs magnitude \\
$absMag_z$	& real	&  4	&  magnitudes	&    rest frame z' abs magnitude \\
 \enddata
\label{tab:param.tbl}

\tablecomments{The parameters contained in the $Photoz$ Table of the
SDSS Science Archive \myurl{http://skyserver.sdss.org/}. See text for
more details.  }
\end{deluxetable}

\end{document}